# Flexible Robust Optimal Bidding of Renewable Virtual Power Plants in Sequential Markets

Hadi Nemati*, Pedro Sánchez-Martín, Álvaro Ortega

Lukas Sigrist, Enrique Lobato, Luis Rouco

Institute for Research in Technology, ICAI, Comillas Pontifical University, Madrid, Spain

*Corresponding author

E-mail address: hnemati@comillas.edu

*Abstract*— **In this paper, a novel approach to define the optimal bidding of renewable-only virtual power plants (RVPPs) in the day-ahead, secondary reserve, and intra-day markets is proposed. To this aim, a robust optimization algorithm is developed to account for the asymmetric nature of the uncertainties that characterize the market prices, as well as the energy production of the RVPP stochastic sources and flexible demand consumption. Simulation results show increased RVPP benefits compared to other existing solutions and demonstrate the potential of renewable sources to further increase their economic competitiveness. The simplicity of the implementation, the computational efficiency, and the flexible robustness are also verified.**

1. Introduction

   *1.1. Motivation*

A Virtual Power Plant (VPP) combines and operates several independent assets in a coordinated manner to provide a flexible and economical solution to utilities that otherwise could not be possible. This combination and coordination can partially mitigate some of the issues related to Non-dispatchable Renewable Energy Sources (ND-RESs), such as their inherent variability and relatively small size compared to conventional fossil-fuel plants [1]. In this regard, different organizations have manifested their interest in increasing the performance of an integrated portfolio of Renewable Energy Sources (RESs) to operate together as a Renewable-only VPP (RVPP), capable of providing flexibility and ancillary services to the electricity markets while being competitive against other market participants (see, e.g., [2]).

The main limitation of RVPPs is the wide set of uncertainties that characterize the behavior of the units they comprise. These include the ND-RES production (from wind speed and solar irradiation), as well as demand consumption patterns. Such uncertainties can significantly affect Day-Ahead (DAM) and/or ancillary service markets (such as the Secondary Reserve Market – SRM) participation of RVPPs and, thus, their expected benefits. The RVPP operator thus needs to take advantage of other market mechanisms, such as Intra-Day Markets (IDMs), to compensate for any forecast errors. On the other hand, the forecasted market prices, used to plan the participation of the RVPP in different markets, are also subject to uncertainties that depend on, e.g., the markets' time scale and forecast horizon [3]. Therefore, handling simultaneously multiple sources of uncertainty with different variabilities is important for RVPP operators to be competitive compared to other market participants [4], [5].

*1.2. Literature Review*

Short-term electricity trading usually spans a time window of 24 hours, and different pools take place prior to the power delivery. According to the quantity traded, short-term electricity markets can also be categorized into energy markets, in which the market operator gives/receives payments according to the amount of energy supplied to/consumed from the network, and ancillary service markets, related to the reliability and security of the grid [6]. In the literature, there is a vast body of papers that model the behavior of VPP in the energy markets, which mainly includes DAM and IDM [7]–[11]. In [7], a risk-constrained Stochastic Programming (SP) method is implemented to capture the uncertainty associated with both the short-term operation of VPP in the energy market and long-term VPP investment decisions. An SP approach is proposed in [8] to model a VPP, including thermal, hydroelectric power plants, and wind units, in the DAM. The uncertainty of wind generation is modeled by defining risk-averse and risk-neutral formulations using conditional value at risk. A decomposition approach is used to increase the computational performance of the optimization problem. In [9], a robust stochastic algorithm is proposed to model the uncertainties of source-side (including wind and solar photovoltaic (PV) plants) and load-side (including different electric, thermal, cooling, and natural gas loads) in the energy market. The uncertainties of wind and PV units' output power are captured with a robust adjustable approach. Load-side uncertainties are modeled by Wasserstein generative adversarial network with gradient penalty. In [10], a multi-objective optimization approach is proposed to model a VPP, including the wind unit, PV station, controllable load, gas turbine, electric vehicle, and gas storage connected to the power-to-gas network. The payoff table is defined to convert the VPP profit and risk objectives to a single-objective form. The uncertainty of wind and PV units are modeled by conditional value at risk and Robust Optimization (RO) method. In [11], the bidding strategy of price taker wind park VPP in the DAM and IDMs of the Spanish market is investigated by a multi-stage SO approach. A stochastic dual dynamic algorithm is used to solve the Markov decision process, defined for the VPP bidding problem. In the above papers, VPP participation in different electricity markets, which can bring more benefit for VPP, is neglected. Besides, the uncertainties in both ND-RESs and electricity prices are barely studied together in their optimization models.

Apart from DAM and IDM, ancillary service markets (ASMs) exist, efficiently assigning resources to guarantee a reliable power system operation. Frequency and non-frequency-related ancillary services exist. Most of such markets are based on the availability of certain levels of power reserves (capacities) that were adequately scheduled in advance. The papers [12]–[16] study the VPP participation in multi-markets, including energy and reserve markets. In [12], a co-optimization method is proposed to model an urban VPP participation in the energy and reserve services considering the Balancing Market (BAM). In [13], a price-taker model is proposed for VPP participation in the energy, reserve, and reactive power markets. The VPP in this paper includes conventional power plants, Energy Storage Systems (ESSs), and interruptible loads. A non-linear model is proposed in [14] for the participation of price-taker VPP in the energy and spinning reserve markets. In the above papers, the authors use deterministic models and neglect the uncertainties. Paper [15] proposes a multi-objective RO problem in order to maximize the operation benefit of VPP and to minimize operation risk and carbon emissions. The authors use weight coefficients and a payoff table to transform the multi-objective problem into a single-objective problem and use fuzzy logic to solve the final Mixed Integer Linear Programming (MILP) problem. In [16], the spatial and temporal uncertainty budgets are defined separately to constrain wind production in terms of time and wind unit numbers. Although VPP offering in simultaneous energy and reserve markets is implemented in the above papers, the complete formulation for reserve provision of each VPP's units and associated constraints are simplified. Besides, the reviewed literature only considers the first sequence (DAM offer + reserve estimations) of a multi-market offering sequence of a VPP. Therefore, a complete model is needed to assess the interactions and sequence of different markets.

Uncertainty modeling is one of the most challenging parts of power system operation and planning [3], [17]. As observed from the discussion given above, two main approaches can be generally followed to model market price and stochastic energy production/consumption in optimization problems, namely SP and RO [18]. In the SP, scenarios are constructed based on the probability distribution of uncertain parameters. The scenarios are then used to configure the optimization problem, in which the final results are assigned according to the probability of the scenarios [18]. In [19], a multi-objective model is implemented for VPP, including wind units, hydroelectric power plants, and thermal units, participation in the continuous IDMs. The uncertainties of hydro inflow forecast, wind generation, and shared order book of VPP compared to other market participants are captured through scenarios by a multi-stage SP model. In [20], [21], a two-stage SP approach is proposed to find the optimal bidding strategy of a VPP in the joint energy and reserve market. The uncertainties of ND-RESs generation, demands, reserve deployment requests, and electricity prices (DAM, spinning reserve market, and BAM) are captured through a scenario generation method. However, constructing a priori scenarios is not always simple or even accurate enough. Moreover, as the number of scenarios increases, the number of constraints and variables in the optimization problem grows, resulting in an exponential increase in computational burden.

By contrast, RO establishes a formulation that allows obtaining the values of the decision variables based on *adverse* realizations of the model parameters. RO has the advantage that the mathematical formulation does not grow in the number of variables and constraints as much as it happens with SP. However, RO results can be too conservative, although the user may define the level of conservatism by selecting the so-called *uncertainty budget* to alleviate this issue [22]. In [23], the RO problem of a price-taker VPP, including wind power plant, demand, and ESS, is proposed in the DAM and BAM. The uncertainties considered include market prices and wind production. In [24], an RO approach is discussed to operate a VPP, including hydro-pumped storage units, thermal units, wind and PV plants, and demand in the energy and reserve markets. Symmetric uncertainties of RESs and demand are considered through confidence bounds (price uncertainty is not considered). The up/down reserve bidding by hydro-pumped storage units and thermal units is modeled in the optimization problem. An iterative algorithm based on solving two MILP problems related to hydro-pumped and thermal units operation is offered to solve the resulting two-stage problem. In [23] and [24], the uncertainty budget is set on an hourly basis for each uncertainty in the constraints, which implies the decision from the VPP operator of dozens of parameters per uncertain profile in each market session.

There are more advanced models to capture the uncertain parameters in which the two approaches (SP and RO) are combined or improved. For example, in the Stochastic Adaptive RO (SARO) approach, the problem is defined as a max-min-max problem [25], [26]. The third level allows modeling the corrective actions after uncertain parameters occur. Therefore, compared to a simple RO model, there is more flexibility to counter uncertainties. The main limitation of the SARO approach is that not many stochasticities can be included in the model, as otherwise, the problem can easily become too cumbersome computationally. SARO approaches are best suited for problems where feasibility must be strictly guaranteed in all ranges of uncertainties considered. In the problem that is solved in this paper, it can safely be assumed that not complying with the RVPP optimal bid can be admitted, albeit some penalization may be applied.

Distributed RO (DRO) technique is a powerful tool, especially when uncertainties are centered around the type of density distribution. Its primary strength lies in scenarios where the exact distribution of the uncertainty is unknown, and one needs to optimize against the worst-case distribution within a certain ambiguity set. The papers [27]–[29] use DRO technique to model the uncertainties of RESs in the DAM. The paper [27] proposes a dispatch strategy for a VPP, including flexible ramping products. A DRO technique is implemented to consider wind power generation uncertainty. The authors in [28] use the DRO approach to take into account the uncertainties of wind power generation and demand in the DAM unit commitment problem. The moment information of the unknown, uncertain parameters is adopted to model the ambiguity set of uncertain parameters. In reference [29], an ambiguity set based on the Wasserstein metric and moment information is utilized to capture the uncertainties of RESs in a combined heat power plant and VPP in the DAM and BAM. In [30], a DRO approach is proposed to model the bidding problem of a VPP in the DAM and BAM. A Wasserstein ambiguity set is used to deal with both uncertainties of electricity price and wind power generation. It is shown that the proposed DRO approach has better out-of-sample performance compared to the SP problem. In the context of this paper, the parameters of the price density functions are not just assumed to be known but are also derived from historical data, making the application of DRO less pertinent.

Non-parametric Probabilistic Modeling is also a potent tool in cases with uncertainties where the underlying distributions are unknown or complex. In [31], an SP tool based on deep learning is proposed for a retailer (also referred to as a VPP in the paper) participating in the energy market. A non-parametric model of prediction errors is used to predict the specified distribution quantiles of uncertainties in electricity prices and ND-RES production. In [32], a probabilistic forecast of ND-RES production is provided for a combined wind/PV VPP bidding in the reserve market. A non-parametric copula-based model captures the non-linear dependency between ND-RESs production and electricity prices. An SRO approach is adopted in [33] to model the VPP in the energy and reserve market. A non-parametric Bayesian model extracts information on different uncertainties related to ND-RES production, demands, and electricity market prices. However, the structure of the sequential electric markets allows for modeling uncertainties using well-established parameters such as means, medians, and quartiles. While non-parametric methods can be employed for data preprocessing, the proposed approach directly integrates these known parameters into the optimization stage, ensuring a more accurate representation.

Regardless of the approach applied, a feature that the aforementioned RO-based references share is that they consider symmetric distributions of the uncertain profiles. This assumption, which simplifies the problem implementation and input data generation, does not accurately represent the reality of, e.g., solar irradiation or wind speed profiles, nor the different market prices. Moreover, regardless of the uncertainty approach applied, the references above consider either conventional (thermal) units and/or fast ESSs (e.g., batteries) as part of the VPP. In other words, the ND-RESs rely on thermal or electrochemical units to improve their reliability. The unique challenges faced by renewable-only RVPPs have not been adequately addressed. Besides, the literature mainly focuses on the first market window (generally the DAM with SRM) due to its larger liquidity. However, how the results of that optimization problem affect subsequent market sessions is not analyzed.

*1.3. Contributions*

The contributions of this paper are thus threefold:

- *A single-level robust optimization model tailored for renewable-only RVPPs to define their optimal bidding in sequential energy and reserve markets:* This model considers the interactions and sequences of different energy and reserve markets, thus being specifically tailored to address the intricacies and challenges posed by sequential markets in the European context. Moreover, the high computational efficiency of the proposed single-level optimization problem allows evaluating, through parametric sensitivity analysis, the impact of choosing different uncertainty budgets for the uncertain parameters in the model, providing the user (e.g., the RVPP operator) with the adequate solution to each market condition. To the authors' knowledge, this type of analysis has not been presented in the context of optimal RVPP bidding.

- *Asymmetric uncertainty on robust optimization:* Existing models assume symmetric uncertainties, which may not accurately represent the real-world scenarios RVPPs face. The proposed robust model handles the asymmetric behavior of uncertain parameters, providing a more realistic representation. An evaluation of how this asymmetric handling impacts RVPP profit is offered. Such evaluation is compared against previously proposed solutions that consider symmetric uncertainty distributions, showcasing the significant differences between the proposed more realistic approach, and the symmetric approximations. The proposed model is also agnostic with respect to the methodology used to generate the forecasts of the probability distributions of the uncertainties. This flexibility ensures that the proposed model remains relevant and adaptable to a wide range of forecasting methodologies, enhancing its practical applicability.

- *Robustness budget on global scheduling horizons:* Previously proposed models often require the definition of uncertainty budgets on an hourly basis for each uncertain parameter in the constraints, which can be cumbersome and less intuitive for RVPP operators. The proposed approach simplifies this by allowing operators to define a single uncertainty budget for each parameter for the entire period. This not only reduces the complexity but also enhances the practicality of the proposed model for real-world applications.

*1.4. Paper Organization*

The remainder of the paper is organized as follows. The RVPP bidding in sequential energy and reserve markets is introduced in Section 2. An overview of the proposed flexible robust RVPP bidding optimization is presented in Section 3, whereas the detailed formulation is provided in Section 4. Section 5 discusses simulation results for different case studies. Finally, conclusions are drawn in Section 6.

## 2. RVPP Bidding in Sequential Energy and Reserve Markets

In this section, the concept of *sequential electricity markets* is introduced, and the bidding strategy of RVPP used in this paper for sequential markets participation is presented. While many system operators opt for joint market clearing, especially in day-ahead and intraday trading, there exist notable instances and specific segments where a sequential market clearing approach is observed. This is particularly evident in the balancing and reserve markets, where individual countries often manage these aspects separately from the main energy market. [18], [34]. This distinction is crucial as the dynamics, rules, and challenges of sequential markets differ from those of joint clearing markets. Further, it is important to distinguish between joint market clearing, joint decision-making for determining optimal bids, and joint bidding. In a sequential market clearing structure, decision-making for determining optimal bids can be made jointly (i.e., different markets can be considered simultaneously at the time of making the optimal decision for the market participant, although such markets are cleared sequentially by the operator). The bids for the different markets can then be submitted jointly or sequentially according to market gate closing times. Hence, the choice is to consider sequentially cleared markets for our RO-based formulation of the bidding model. Figure 1 depicts the Spanish electricity market as an example of a typical market sequence in European countries [35]. The Spanish energy markets include DAM and seven IDMs. The generation units and demands send DAM offers for an upcoming day, including both electricity price and energy, to the Spanish Market Operator (OMIE). Units have opportunities to adjust their DAM energy offers by participating in IDMs. The sequential optimization problems are solved in this paper based on the fact that more accurate forecasts of uncertain parameters could be available as time is closer to the power delivery period.

In addition to energy markets, several reserve markets are run to guarantee the security of the system. The SRM associated with Automatic Generation Control (AGC) is run by the Spanish Transmission System operator (TSO), Red Eléctrica de España (REE). In this regard, the resources of the up and down-regulation provision for the 24 hours of the operation day are assigned by REE for seven operational zones of the Spanish system. Then, each zone is responsible for providing its promised reserve by clearing the offers from its reserve providers. Note that the electric market design shown in Figure 1 can be adapted, upon some adjustments, to the situation in other European countries.

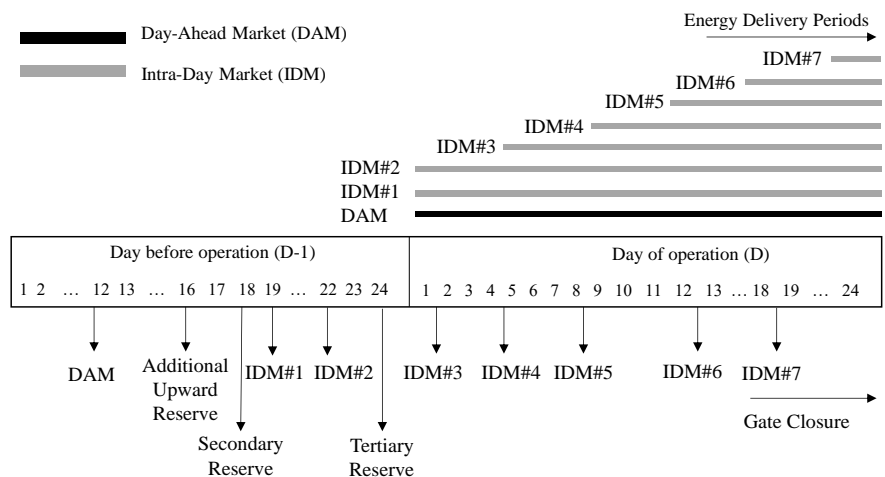

Figure 1 Spanish electricity markets structure [35].

Figure 2 shows the bidding strategy of RVPP in the sequential markets, including DAM, SRM, and IDMs. The information flow between different electricity markets and the necessary input data for running the optimization problems in each of the mentioned electricity markets are also presented. In this paper a price-taker RVPP, i.e., an RVPP with *small* size, is modeled. Therefore, the RVPP does not significantly affect the market clearing price. As market prices are unknown until the market is cleared, the RVPP optimization model considers electricity price forecasts data as inputs, which are generated by means of large sets of historical data. Note that the problem is solved from the RVPP operator point of view as a bidding unit, not from the overall system operator's side. In other words, the RVPP operator does not have access to the bids submitted by other market participants. Therefore, the dual form of the network operation problem is not considered in order to clear the market and to determine the nodal electricity price. Instead, the worst cases of income deviations in the corresponding bounds of electricity price forecast (as well as the forecast bounds of ND-RESs energy and demands) are obtained in the robust optimization problem. Moreover, it is assumed that RVPP bids at zero or low prices. This assumption is relevant since the size of RVPP is small, and it mainly includes ND-RESs with low operation costs. Note that the proposed bidding approach could potentially be adapted for joint clearing markets. For example, if DAM and SRM are cleared at the same time, the RVPP can use the first pane of Figure 2 to bid in these two markets at the same time.

According to Figure 2, the step-by-step solution methodology of the optimization problems for different market sessions in the Spanish Market is summarized below:

1. The RVPP receives the input data of its units in addition to the forecast of uncertain parameters before the gate closure of each electricity market selected for participation.

2. The RVPP maximizes its benefits by solving different single-level MILP optimization problems defined in Section 4 related to each market session.

3. The results of solving the optimization problems are used to bid energy or reserve in the corresponding electricity market and to determine the dispatch of RVPP units.

4. The bids submitted by RVPP to the markets are cleared based on the priorities of OMIE and REE, and then the results are sent to the RVPP operator. The energy and reserve bids may or may not be accepted according to the decision of the market operator. Note that the RVPP finds the optimal bids according to the forecast data and whether they are feasible or not will be determined by the system operator.

5. The RVPP uses its accepted bids from previous market sessions as well as the updated input data and forecast of uncertain parameters to solve the optimization problem for the upcoming market session.

Note that in the proposed RVPP bidding scheme, the interconnection between the markets is appropriately addressed. For example, in the case of the DAM objective function, the possibility of providing reserve in the SRM needs to be considered. Therefore, the optimization problem for both DAM and SRM is performed (see upper pane of Figure 2). The RVPP offers energy to the DAM before the DAM gate closure. However, the up/down reserve is not offered and can be reoptimized before the SRM gate closure. In addition, the first IDM (IDM#1) effect in the SRM objective function is considered, but the IDM#1 bid results are not offered in the SRM (see middle pane of Figure 2). Finally, the objective function for each IDM is optimized by considering the bidding results of DAM, SRM, and previous IDMs (see bottom pane of Figure 2). The proposed bidding approach allows the RVPP to omit some of the above electricity markets if so deemed. For instance, the SRM or a number of IDMs can be ignored, and RVPP can go for the subsequent market, in which the input data is based on previously implemented markets. It is worth noting that IDMs are not included in the co-optimization problem of the DAM snapshot. The reason is that the co-optimization of DAM and IDMs can lead to a speculation effect. For instance, a net seller RVPP might attempt to benefit by, e.g., buying a large amount of energy in the DAM to sell it later at a higher price in the IDMs or vice versa. However, this co-optimization is not implemented because usually, the regulator of the system does not allow this kind of participation in the market. Besides, IDMs tend to have much less liquidity than DAM and are mainly used for minor energy adjustments (however, there are some IDM markets with relatively high traded energy such as IDM#1 in the Spanish system). Therefore, with low market liquidity, there is a high risk of not being called in the IDM, thus potentially incurring notable economic losses for the RVPP.

The optimization problems involved in the different electricity markets shown in Figure 2, namely (i) DAM+SRM; (ii) SRM+IDM#1; and (iii) IDM#k, will be developed in Sections 3 and 4 of this paper.

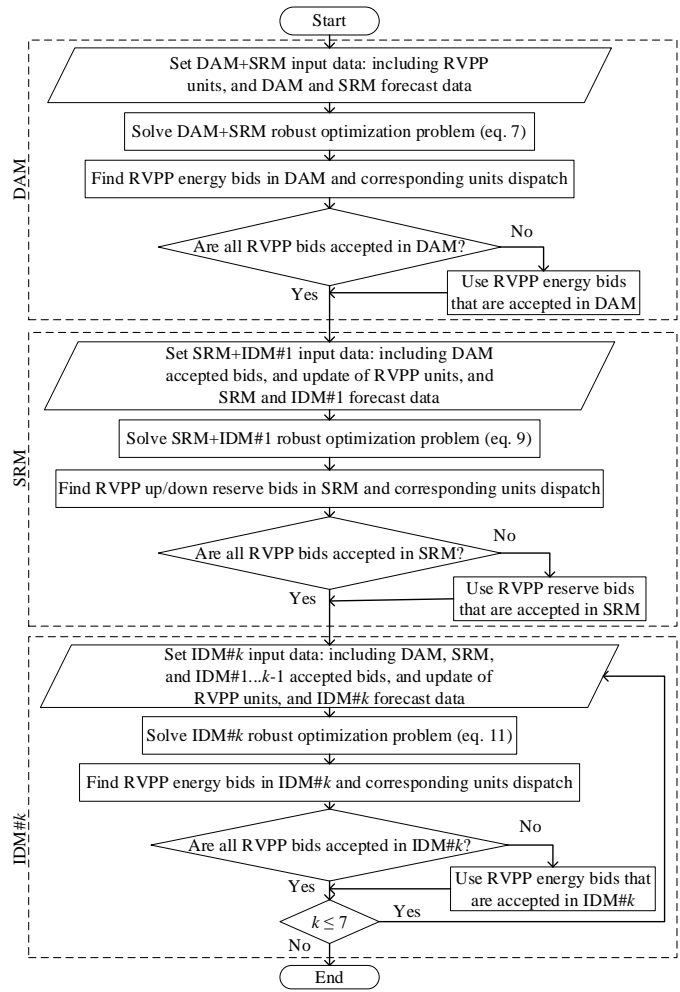

Figure 2 Bidding strategy of RVPP in the sequential electricity markets.

## 3. Overview of Flexible Robust RVPP Bidding Optimization

The majority of the implementations of flexible RO in VPP applications (e.g., [1], [9], [26], [36]) are based on the principles that Bertsimas and Sim presented in [22]. In the following sections, the approach of Bertsimas and Sim, which is summarized in Appendix A for interested readers, is further developed to consider the asymmetric uncertainty in the objective function and to define the robustness budget on global scheduling horizons of the associated constraints.

### 3.1. Flexible Robust Optimization of the Objective Function

In this section, the flexible robust formulation of the objective function is formulated by developing the formulation in [22]. For this purpose, consider the following linear deterministic problem:

$$max \sum_t C_t x_t \quad (1a)$$

s.t.:

$$Ax = B \quad (1b)$$

$$C \leq x \leq D \quad (1c)$$

where $C_t$ is the vector of coefficients of objective function (1a); $x_t$ is the vector of free decision variables; $t$ is the time index representing hours; $A$ and $B$ are matrices of fixed parameters in the equality constraint (1b); $C$ and $D$ are the vectors of lower and upper bounds of inequality constraint (1c), respectively.

To develop the flexible robust optimization on the objective function, let us assume the parameter $C_t$ as a bounded uncertain parameter with the symmetric probability distribution at time $t$ (i.e., $C_t \in [\widetilde{C}_t - \widehat{C}_t, \widetilde{C}_t + \widehat{C}_t]$), where $\widetilde{C}_t$ is the mean of the uncertain distribution, and $\widehat{C}_t$ is the deviation at time $t$. By elaborating on the approach followed in [22] (see Appendix A), (2a)-(2f) is presented as the robust formulation of (1a)-(1c):

$$max \sum_t \widetilde{C}_t x_t - \Gamma^o v^o - \sum_t \eta_t^o \quad (2a)$$

s.t.:

$$v^o + \eta_t^o \geq \widehat{C}_t y_t^o \quad \forall t \quad (2b)$$

$$-y_t^o \leq x_t \leq y_t^o \quad \forall t \quad (2c)$$

$$v^o, \eta_t^o, y_t^o \geq 0 \quad \forall t \quad (2d)$$

$$Ax = B \quad (2e)$$

$$C \leq x \leq D \quad (2f)$$

where $\Gamma^o$ is a user-defined parameter that represents the robust uncertainty budget in the objective function; $v^o$ and $\eta_t^o$ are dual variables related to the parameter uncertainty; and $y_t^o$ is the auxiliary variable of the $x_t$ absolute value function. The first term in (2a) is like the deterministic objective function (1a) by substituting the mean value of uncertain parameter $C_t$ ($\widetilde{C}_t$). The second and third terms in (2a) calculate the objective function reduction due to uncertainty of parameter $C_t$.

In this paper, the asymmetric behavior of uncertain parameters is modeled. To this aim, let us assume the parameter $C_t$ as a bounded uncertain parameter with the asymmetric probability distribution at time $t$ (i.e., $C_t \in [\widetilde{C}_t - \breve{C}_t, \widetilde{C}_t + \widehat{C}_t]$), where $\breve{C}_t$ is the negative deviation at time $t$; $\widehat{C}_t$ is the positive deviation at time $t$; and $\widetilde{C}_t$ is the median of the distribution (not the mean). To model the asymmetric behavior in the objective function, constraint (2c) is replaced by eq. (3):

$$-\frac{\widehat{C}_t}{\breve{C}_t} y_t^o \leq x_t \leq y_t^o \qquad \forall t \qquad (3)$$

In the above constraint, if $\widehat{C}_t = \breve{C}_t$, then constraint (2c) and (3) are formally equivalent.

As can be observed from (2a)-(2f), the uncertainty budget is implemented using a parameter $\varGamma^o$ that adjusts the level of robustness that the user chooses for each source of uncertainty. This uncertainty budget $\varGamma^o$ represents the number of time periods over the whole scheduling horizon for which an uncertain parameter will take the value that has the worst impact on the objective function. For instance, for a net seller RVPP, prices will take the lowest value in their interval for $\varGamma^o$ periods in (2a)-(2f), whereas for a net buyer RVPP, the opposite applies.

The uncertainty budget $\varGamma^o$ can thus take any value between zero (deterministic case without uncertainty – optimistic scenario) and the number of periods of the market horizon, $T$ (pessimistic scenario). For instance, for the DAM, uncertain prices are defined for 24 periods of 1 hour. If $\varGamma^o$ takes any value between 0 and 24, then the optimization algorithm will choose the $\varGamma^o$ periods for which the worst realization of the uncertain parameter would impact the most on the objective function. For the remaining $24 - \varGamma^o$ periods, the uncertain parameter would be deterministic, i.e., $C_t = \widetilde{C}_t$. If non-integer values for $\varGamma^o$ are chosen, a fraction of the last period in which uncertainty is included would be considered. With this formulation, the user can select the strategy to follow, either optimistic (low values of $\varGamma^o$) or conservative/pessimistic (high values of $\varGamma^o$).

### 3.2. Flexible Robust Optimization of Time-varying Constraints

The uncertainty of stochastic renewable production and demand consumption also needs to be considered in the RVPP optimization problem. However, these uncertain parameters appear in the optimization constraints, which further convolutes the problem and its formulation. To model the flexible robust constraints for stochastic renewable production, let us assume the following deterministic inequality constraint:

$$x_t \leq X_t \qquad \forall t \qquad (4)$$

where $X_t$ is the vector of positive upper bounds of inequality constraint (4).

To develop the flexible robust optimization of the sequential constraints for renewable productions, let us assume parameter $X_t$, which represents the available power production of a given ND-RES, as a bounded uncertain parameter which can take values within the interval $X_t \in [\widetilde{X}_t - \breve{X}_t, \widetilde{X}_t]$, in which $\widetilde{X}_t$ is the median of the uncertain distribution, and $\breve{X}_t$ is the negative deviation at time $t$. As uncertainty is affecting the ND-RES generation, the scenario that shows the most negative impact on the objective function is that for which, for a given period, the uncertain parameter takes the lowest value of its interval, i.e. $X_t = \widetilde{X}_t - \breve{X}_t$. Thus, the positive deviation of ND-RESs available power, $\widehat{X}_t$, is not considered in the proposed approach. The opposite applies in the case of demands, i.e., the downward deviation of demand consumption is not considered. By elaborating on the approach followed in [9], the robust formulation (5a)-(5h) is proposed to consider the uncertain parameter $X_t$ in the constraint (4):

$$x_t \leq \widetilde{X}_t - y_t^c \quad \forall t \quad (5a)$$

$$y_t^c \leq \overline{X}_t \quad \forall t \quad (5b)$$

$$y_t^c \geq v^c + \eta_t^c - M(1 - \chi_t) \quad \forall t \quad (5c)$$

$$v^c + \eta_t^c \geq \overline{X}_t \quad \forall t \quad (5d)$$

$$\varepsilon \chi_t \leq \eta_t^c \leq M \chi_t \quad \forall t \quad (5e)$$

$$\sum_t \chi_t = \Gamma^c \quad (5f)$$

$$v^c, \eta_t^c, y_t^c \geq 0 \quad \forall t \quad (5g)$$

$$\chi_t \in \{0,1\} \quad \forall t \quad (5h)$$

where $\Gamma^c$ is a user-defined parameter that represents the robust uncertainty budget in the optimization constraints; $v^c$ and $\eta_t^c$ are dual variables related to the sensitivity of the upper bound value to the robust uncertainty budget; $y_t^c$ represents the negative deviation of the uncertain parameter; $\varepsilon/M$ is a very small/big positive number; and $\chi_t$ is a binary variable that guarantees the robustness budget predefined by the user through $\Gamma^c$ ($\chi_t = 1$, if for period $t$ the uncertain parameter deviates with respect to the median value).

To develop the flexible robust constraints for demands, parameter $X_t$ is assumed as a bounded uncertain parameter with values within $X_t \in [\widetilde{X}_t, \widetilde{X}_t + \widehat{X}_t]$, in which $\widetilde{X}_t$ is the median of the uncertain distribution, and $\widehat{X}_t$ is the positive deviation at time $t$. Then, constraints (5a), (5b), and (5d) are replaced by (6a), (6b), and (6c), respectively:

$$x_t \leq \widetilde{X}_t + y_t^c \quad \forall t \quad (6a)$$

$$y_t^c \leq \widehat{X}_t \quad \forall t \quad (6b)$$

$$v^c + \eta_t^c \geq \widehat{X}_t \quad \forall t \quad (6c)$$

$$(5c), (5e)-(5h) \quad (6d)$$

The set of constraints (5a)-(5h) (and accordingly (6a)-(6d)) needs to be defined for each source of uncertainty in the constraints, i.e., one for each ND-RES and flexible demand unit $i$ that the RVPP contains. This implies that the RVPP operator needs to define as many $\Gamma_i^c$ as ND-RESs and flexible demands are included in the RVPP. As opposed to price uncertainty budget $\Gamma^o$, each $\Gamma_i^c$ can only take integer values between 0 and the number of periods of each market. This is due to the binary nature of $\chi_t$ in (5a)-(5h), which does not allow for non-integer values for $\Gamma_i^c$. Implementing a robustness budget on the global scheduling horizons, the higher $\Gamma_i^c$ is, the more periods in the scheduling horizon include this uncertainty in the model, and thus, the more conservative the solution is. The power deviations of different time periods are chosen based on their magnitude following a decreasing path. That means those time periods that lead to the highest deviation of ND-RES production or demand are chosen before other periods. It is worth mentioning that, in contrast to the proposed model, the literature (e.g., [1], [23]) proposes defining a separate uncertainty budget for each time period of each ND-RES or demand. In the references above, the operator needs to assign substantially more uncertainty budgets for each of the units, which, in most cases, is not an easy task. Besides, it can be inferred that by increasing the value of the uncertain parameter for each time period, the uncertainty evenly affects the objective function instead of selecting the worst scenarios.

Finally, the feasibility of the optimization problem is related to the value of parameter $\Gamma_i^c$. For the maximum value of $\Gamma_i^c$, the optimization problem is feasible for all possible deviations of uncertain parameters in its predicted bounds. For lower values of $\Gamma_i^c$, the optimization problem is feasible for at least the number of worst periods in which energy deviates. The deviation of energy in other periods leads to the probabilistic feasibility of the solution. That means as energy deviates in more time periods with a more considerable amount, the feasibility of the optimization problem decreases.

Note that the RVPP operator bases optimal bid decisions on forecast data, leaving the actual feasibility assessment of these bids (and of other market participants) to the system operator, which is beyond the scope of this paper. In this paper, we introduce in the case studies of Section 5 the concept of "possible unfeasibility" to assess the RVPP's ability to meet market bids considering potential energy fluctuations, since our model optimizes bids within forecast uncertainty limits without pre-judging their feasibility.

The following section builds upon the concepts of robust optimization discussed above, and provides with the detailed formulation proposed in this paper to determine the optimal bidding of RVPPs in sequential energy and reserve markets.

## 4. Flexible Robust Optimization Model

This section presents and describes the robust formulation of the RVPP bidding problem in the sequential electricity markets, based on the concepts on robust optimization outlined in Section 3. The nomenclature used in the remainder of this Section is first detailed in Section 4.1. The electricity price uncertainty modeling in the objective function for each DAM+SRM, SRM+IDM#1, and IDM#k problem is then presented in Section 4.2. Section 4.3 formulates the constraints that do not include any uncertain parameters, such as the supply-demand balancing constraints and the power traded constraints. Finally, the uncertainty modeling in the constraints of ND-RES production and flexible demand consumption is discussed in Section 4.4.

### 4.1. Nomenclature

The notation considered in the proposed RVPP model is presented below. The acronym STU used in the definition of certain parameters/variables stands for Solar Thermal Units.

#### 4.1.1. Indexes and Sets

| Symbol | Description |
|---|---|
| $d \in D$ | Set of demands |

| Symbol | Description |
|---|---|
| $k \in K$ | Set of IDM sessions |
| $p \in P$ | Set of daily load profiles |
| $r \in R$ | Set of ND-RESs |
| $t \in T / t \geq \tau$ | Set of time periods/IDM time periods |
| $\theta \in \Theta$ | Set of STUs |
| $\Xi^{DA}/\Xi^{ID}/\Xi^{SR}$ | Set of decision variables of DAM/IDMs/SRM |

### 4.1.2. Parameters

| Symbol | Description |
|---|---|
| $C_{d,p}$ | Cost of load profile $p$ of demand $d$ [€] |
| $C_r^R$ | Operation and maintenance costs of ND-RES $r$ [€/MWh] |
| $\underline{E}_d$ | Minimum energy consumption of demand $d$ throughout the planning horizon [MWh] |
| $M$ | Very big positive value [-] |
| $\widetilde{P}_{d,p,t}^{DA}/\widehat{P}_{d,p,t}^{DA}$ | Median/positive deviation of hourly consumption of profile $p$ of demand $d$ during period $t$ [MW] |
| $\underline{P}_d/\overline{P}_d$ | Minimum/maximum power consumption of demand $d$ [MW] |
| $\underline{P}_r/\overline{P}_r$ | Minimum/maximum power production of ND-RES $r$ [MW] |
| $\underline{P}_\theta/\overline{P}_\theta$ | Minimum/maximum power production of STU $\theta$ [MW] |
| $\widetilde{P}_{r(\theta),t}^{DA}/\widecheck{P}_{r(\theta),t}^{DA}$ | Median/negative deviation of ND-RES $r$ (solar field of STU $\theta$ thermal) production in the DAM during period $t$ [MW] |
| $\widetilde{P}_{r,t}^{SR}/\widecheck{P}_{r,t}^{SR}$ | Median/negative deviation of ND-RES $r$ production in the SRM during period $t$ [MW] |
| $\widetilde{P}_{k,r,t}^{ID}/\widecheck{P}_{k,r,t}^{ID}$ | Median/negative deviation of ND-RES $r$ production in the IDM#$k$ during period $t$ [MW] |
| $\underline{R}_d/\overline{R}_d$ | Down/up ramp rate of demand $d$ [MW/hour] |
| $\underline{R}_{r(d)}^{SR}/\overline{R}_{r(d)}^{SR}$ | Secondary Reserve (SR) down/up ramp rate of ND-RES $r$ (demand $d$) [MW/min] |
| $T^{SR}$ | Required time for SR action [min] |
| $\underline{\beta}_{d,t}/\overline{\beta}_{d,t}$ | The percentage of down/up flexibility of demand during period $t$ [%] |
| $\Gamma^{DA}/\Gamma_k^{ID}$ | DAM/IDM#$k$ price uncertainty budget [-] |
| $\Gamma^{SR,\uparrow}/\Gamma^{SR,\downarrow}$ | Up/down SRM price uncertainty budget [-] |
| $\Gamma_{r(\theta)}^{DA}$ | ND-RES $r$ (Solar field of STU $\theta$ thermal) production uncertainty budget in the DAM [-] |
| $\Delta t$ | Duration of periods [hour] |
| $\varepsilon$ | Very small positive value [-] |
| $\kappa$ | User-defined parameter to assign the limit of up reserve traded in the market according to a share of the total power capacity of the RVPP [-] |
| $\widetilde{\lambda}_t^{DA}/\widehat{\lambda}_t^{DA}/\widecheck{\lambda}_t^{DA}$ | Median and positive/negative deviation of DAM price prediction during period $t$ [€/MWh] |

| Symbol | Description |
|---|---|
| $\tilde{\lambda}_t^{SR,\uparrow}/\hat{\lambda}_t^{SR,\uparrow}/\check{\lambda}_t^{SR,\uparrow}$ | Median and positive/negative deviation of up SRM price prediction during period $t$ [€/MW] |
| $\tilde{\lambda}_t^{SR,\downarrow}/\hat{\lambda}_t^{SR,\downarrow}/\check{\lambda}_t^{SR,\downarrow}$ | Median and positive/negative deviation of down SRM price prediction during period $t$ [€/MW] |
| $\tilde{\lambda}_{k,t}^{ID}/\hat{\lambda}_{k,t}^{ID}/\check{\lambda}_{k,t}^{ID}$ | Median and positive/negative deviation of IDM#$k$ price prediction during period $t$ [€/MWh] |
| $\varrho_t$ | A coefficient to calculate the proportion of requested down reserve to up reserve by the TSO during period $t$ [-] |

### 4.1.3. Continuous Variables

| Symbol | Description |
|---|---|
| $p_{d,t}^{DA}$ | Consumption of demand $d$ in the DAM during period $t$ [MW] |
| $p_{r,t}^{DA}$ | Production of ND-RES $r$ in the DAM during period $t$ [MW] |
| $p_{\theta,t}^{DA}$ | Production of STU $\theta$ in the DAM during period $t$ [MW] |
| $p_{k,d,t}^{ID}$ | Consumption of demand $d$ in the IDM#$k$ during period $t$ [MW] |
| $p_{k,r,t}^{ID}$ | Production of ND-RES $r$ in the IDM#$k$ during period $t$ [MW] |
| $p_{k,\theta,t}^{ID}$ | Production of STU $\theta$ in the IDM#$k$ during period $t$ [MW] |
| $p_t^{DA}/p_{k,t}^{ID}$ | Total traded power by RVPP in the DAM/IDM#$k$ during period $t$ [MW] |
| $p_{\theta,t}^{SF}$ | Thermal power output of the solar field of STU $\theta$ during period $t$ [MW] |
| $r_t^{SR,\uparrow}/r_t^{SR,\downarrow}$ | Total up/down SR traded by RVPP during period $t$ [MW] |
| $r_t^{SR}$ | Total SR traded by RVPP for different TSO calls on conditions during period $t$ [MW] |
| $r_{r(d),t}^{SR,\uparrow}/r_{r(d),t}^{SR,\downarrow}$ | Up/down SR provided by ND-RES $r$ (demand $d$) during period $t$ [MW] |
| $r_{r(d),t}^{SR}$ | SR provided by ND-RES $r$ (demand $d$) for different TSO calls on conditions during period $t$ [MW] |
| $r_{\theta,t}^{SR}$ | SR provided by STU $\theta$ for different TSO calls on conditions during period $t$ [MW] |
| $y_t^{DA}/y_{k,t}^{ID}$ | Production affected by DAM/IDM#$k$ price uncertainty during period $t$ [MWh] |
| $y_{r(\theta),t}^{DA}$ | Variation of ND-RES $r$ (solar field of STU $\theta$ thermal) production during period $t$ [MW] |
| $y_{d,t}^{DA}$ | Variation of demand $d$ consumption during period $t$ [MW] |
| $\eta_t^{DA}/\eta_{k,t}^{ID}$ | Dual variable to model the price uncertainty of DAM/IDM#$k$ during period $t$ [€] |
| $\eta_t^{SR,\uparrow}/\eta_t^{SR,\downarrow}$ | Dual variable to model the price uncertainty of up/down SRM during period $t$ [€] |
| $\eta_{r(\theta),t}^{DA}$ | Dual variable to model the ND-RES $r$ (solar field of STU $\theta$ thermal) production uncertainty during period $t$ [MW] |
| $\eta_{d,t}^{DA}$ | Dual variable to model the demand $d$ uncertainty during period $t$ [MW] |
| $\nu^{DA}/\nu_k^{ID}$ | Dual variable to model the price uncertainty of DAM/IDM#$k$ [€] |
| $\nu^{SR,\uparrow}/\nu^{SR,\downarrow}$ | Dual variable to model the price uncertainty of up/down SRM [€] |
| $\nu_{r(\theta)}^{DA}$ | Dual variable to model the ND-RES $r$ (solar field of STU $\theta$) production uncertainty [MW] |

| $v_d^{DA}$ | Dual variable to model the demand *d* uncertainty [MW] |

### 4.1.4. Binary Variables

| Symbol | Description |
|---|---|
| $u_{d,p}$ | Indicator of selection of profile *p* of demand *d* |
| $\chi_{r(\theta),t}^{DA}$ | Binary variable that is 1 if ND-RES *r* (STU $\theta$) robust constraints are active during period *t*, and 0 otherwise |
| $\chi_{d,t}^{DA}$ | Binary variable that is 1 if demand *d* robust constraints are active during period *t*, and 0 otherwise |

### 4.2. RVPP Flexible Robust Objective Functions

The objective functions of the three problems illustrated in Figure 2, namely DAM (considering possible SRM participation), SRM (jointly with first IDM session), and IDMs, are presented and discussed in Sections 4.2.1, 4.2.2, and 4.2.3, respectively. Constraints that are associated to such objective functions are also included.

#### 4.2.1. DAM +SRM

The objective function (7) maximizes the benefits of RVPP in the DAM and SRM. The first line of (7) calculates the expected RVPP incomes from bidding in the DAM, up SRM, and down SRM. The parameters $\tilde{\lambda}_t^{DA}$, $\tilde{\lambda}_t^{SR,\uparrow}$, and $\tilde{\lambda}_t^{SR,\downarrow}$ are the median of the uncertain parameters (DAM, up SRM, and down SRM prices) in the objective function. The first line of (7) is analogous to the first term of the objective function (2a) in Section 3.1. The second line depicts the operation costs of ND-RESs and the costs of selecting a particular load profile, and it does not include any uncertain parameter. If single deterministic values are considered for parameters $\tilde{\lambda}_t^{DA}$, $\tilde{\lambda}_t^{SR,\uparrow}$, and $\tilde{\lambda}_t^{SR,\downarrow}$, the first and second lines of (7) behave like the deterministic objective function in the DAM and SRM. In the proposed model, the uncertainty of DAM and SRM prices is considered by finding the worst cases of price deviations in the corresponding forecast bounds. The DAM and SRM price deviations due to uncertainty affect the median income from the first line of the objective function (7). The income reduction due to DAM and SRM price uncertainties is determined by the last two lines of the objective function (7). The last two lines of (7) are the reduction of the total expected incomes of the first line of the objective function (7). They represent the worst realization of the price deviations that consequently has the highest impact on reducing the total benefit of the optimization problem. The last two lines in (7) are analogous to the second and third terms of the objective function (2a). The robust variables $v^{DA}$, $v^{SR,\uparrow}$, and $v^{SR,\downarrow}$ represent the average income reduction per uncertain parameter whose deviation is applied to. The robust variables $\eta_t^{DA}$, $\eta_t^{SR,\uparrow}$, and $\eta_t^{SR,\downarrow}$ represent the additional hourly incomes reduction for each price deviation of each market. The uncertainty budgets $\Gamma^{DA}$ and $\Gamma^{SR,\uparrow}/\Gamma^{SR,\downarrow}$ make the robustness of the model flexible against uncertainties in the DAM and SRM electricity prices, respectively.

$$\max_{\Xi^{DA}} \sum_{t \in T} \left( \tilde{\lambda}_t^{DA} p_t^{DA} \Delta t + \tilde{\lambda}_t^{SR,\uparrow} r_t^{SR,\uparrow} + \tilde{\lambda}_t^{SR,\downarrow} r_t^{SR,\downarrow} \right)$$
$$- \sum_{t \in T} \sum_{r \in R} C_r^R p_{r,t}^{DA} \Delta t - \sum_{d \in D} \sum_{p \in P} C_{d,p} u_{d,p} \quad (7)$$
$$- \Gamma^{DA} v^{DA} - \Gamma^{SR,\uparrow} v^{SR,\uparrow} - \Gamma^{SR,\downarrow} v^{SR,\downarrow}$$
$$- \sum_{t \in T} \left( \eta_t^{DA} + \eta_t^{SR,\uparrow} + \eta_t^{SR,\downarrow} \right)$$

Constraints (8a) and (8b) model the impact of DAM price volatility on income reduction when electricity price takes its worst condition value by considering price asymmetry. Depending on RVPP selling or buying electricity in the market, the worst DAM price conditions are, respectively, at the prices values $\tilde{\lambda}_t^{DA} - \check{\lambda}_t^{DA}$ or $\tilde{\lambda}_t^{DA} + \hat{\lambda}_t^{DA}$. Constraints (8c) and (8d) calculate the volatility of up and down SRM income reduction when the reserve price takes its worst value, i.e., $\tilde{\lambda}_t^{SR,\uparrow} - \check{\lambda}_t^{SR,\uparrow}$ and $\tilde{\lambda}_t^{SR,\downarrow} - \check{\lambda}_t^{SR,\downarrow}$. The robust formulation selects the hours for price deviation that affect the most (negatively) the incomes of the objective function. As variables $r_t^{SR,\uparrow}$ and $r_t^{SR,\downarrow}$ are positive, there is no need to evaluate the absolute value of these variables. Constraint (8e) defines the nature of positive auxiliary variables. Constraints (8a)-(8e) are written analogous to (2b), (3), and (2d) in Section 3.1, considering three uncertain parameters related to the DAM, up SRM, and down SRM prices.

$$v^{DA} + \eta_t^{DA} \geq \hat{\lambda}_t^{DA} y_t^{DA} \qquad \forall t \quad (8a)$$

$$-\frac{\hat{\lambda}_t^{DA}}{\check{\lambda}_t^{DA}} y_t^{DA} \leq p_t^{DA} \Delta t \leq y_t^{DA} \qquad \forall t \quad (8b)$$

$$v^{SR,\uparrow} + \eta_t^{SR,\uparrow} \geq \check{\lambda}_t^{SR,\uparrow} r_t^{SR,\uparrow} \qquad \forall t \quad (8c)$$

$$v^{SR,\downarrow} + \eta_t^{SR,\downarrow} \geq \check{\lambda}_t^{SR,\downarrow} r_t^{SR,\downarrow} \qquad \forall t \quad (8d)$$

$$v^{DA}, v^{SR,\uparrow}, v^{SR,\downarrow}, \eta_t^{DA}, \eta_t^{SR,\uparrow}, \eta_t^{SR,\downarrow}, y_t^{DA} \geq 0 \qquad \forall t \quad (8e)$$

*4.2.2. SRM + IDM#1*

The objective function (9) maximizes the benefits of RVPP in the SRM and IDM#1. The deterministic and robust components of the objective function are distinguishable analogously to Section 4.2.1, considering that the uncertain parameters are related to the up SRM, down SRM, and IDM#1 prices. According to the first line of the objective function (9), before the SRM gate closure, the RVPP maximizes the profits of selling up/down reserve in the SRM and energy in IDM#1. Therefore, the possibility of considering the arbitrage opportunity between SRM and IDM#1 is provided. The second line shows the rescheduled operation costs of ND-RES in IDM#1. The profit reduction due to uncertainties in the SRM and IDM#1 electricity prices are considered through the third and fourth lines of (9).

$$\max_{\Xi^{SR}} \sum_{t \in T} \left( \tilde{\lambda}_t^{SR,\uparrow} r_t^{SR,\uparrow} + \tilde{\lambda}_t^{SR,\downarrow} r_t^{SR,\downarrow} + \tilde{\lambda}_{(k=1),t}^{ID} p_{(k=1),t}^{ID} \Delta t \right)$$
$$- \sum_{t \in T} \sum_{r \in R} C_r^R p_{(k=1),r,t}^{ID} \Delta t \qquad (9)$$
$$- \Gamma^{SR,\uparrow} v^{SR,\uparrow} - \Gamma^{SR,\downarrow} v^{SR,\downarrow} - \Gamma_{(k=1)}^{ID} v_{(k=1)}^{ID}$$
$$- \sum_{t \in T} \left( \eta_t^{SR,\uparrow} + \eta_t^{SR,\downarrow} + \eta_{(k=1),t}^{ID} \right)$$

Constraints (10a)-(10d) set the corresponding benefit reductions of the SRM objective function (9). The first two constraints correspond to the SRM, and the last two constraints to IDM#1. To implement the flexible robustness for IDM#1, the asymmetric price deviation should be the one that sets a resulting price of $\tilde{\lambda}_{(k=1),t}^{ID} - \check{\lambda}_{(k=1),t}^{ID}$ when the RVPP is selling energy to IDM#1 and price of $\tilde{\lambda}_{(k=1),t}^{ID} + \hat{\lambda}_{(k=1),t}^{ID}$ for buying energy. Constraint (10e) defines the nature of positive auxiliary variables.

$$v^{SR,\uparrow} + \eta_t^{SR,\uparrow} \geq \check{\lambda}_t^{SR,\uparrow} r_t^{SR,\uparrow} \qquad \forall t \qquad (10a)$$

$$v^{SR,\downarrow} + \eta_t^{SR,\downarrow} \geq \check{\lambda}_t^{SR,\downarrow} r_t^{SR,\downarrow} \qquad \forall t \qquad (10b)$$

$$v_{(k=1)}^{ID} + \eta_{(k=1),t}^{ID} \geq \hat{\lambda}_{(k=1),t}^{ID} y_{(k=1),t}^{ID} \qquad \forall t \qquad (10c)$$

$$-\frac{\hat{\lambda}_{(k=1),t}^{ID}}{\check{\lambda}_{(k=1),t}^{ID}} y_{(k=1),t}^{ID} \leq p_{(k=1),t}^{ID} \Delta t \leq y_{(k=1),t}^{ID} \qquad \forall t \qquad (10d)$$

$$v^{SR,\uparrow}, v^{SR,\downarrow}, v_{(k=1)}^{ID}, \eta_t^{SR,\uparrow}, \eta_t^{SR,\downarrow}, \eta_{(k=1),t}^{ID}, y_{(k=1),t}^{ID} \geq 0 \qquad \forall t \qquad (10e)$$

### 4.2.3. IDM#k

The objective function of IDM#k participation is provided in (11), where the first term calculates the bidding income in each IDM. The second term computes the rescheduled deterministic operation costs of ND-RES in each IDM. The uncertainties in IDMs electricity prices are captured by the third and fourth terms of (11). The robustness of income reduction in the IDMs objective function (11) comes from IDMs price volatility. The uncertainty budget $\Gamma_k^{ID}$ should be assigned according to the different number of periods of each IDM session.

$$\max_{\Xi_k^{ID}} \sum_{t \geq \tau} \tilde{\lambda}_{k,t}^{ID} p_{k,t}^{ID} \Delta t - \sum_{t \geq \tau} \sum_{r \in R} C_r^R p_{k,r,t}^{ID} \Delta t - \Gamma_k^{ID} v_k^{ID} - \sum_{t \geq \tau} \eta_{k,t}^{ID} \qquad \forall k \qquad (11)$$

Constraints (12a)-(12c) calculate the income reduction arising from the worst case of IDM#k electricity price in the IDM objective function. The price deviations will be those that reduce the expected incomes the most, considering the possible asymmetry in price.

$$v_k^{ID} + \eta_{k,t}^{ID} \geq \hat{\lambda}_{k,t}^{ID} y_{k,t}^{ID} \qquad \forall k, t \geq \tau \qquad (12a)$$

$$-\frac{\hat{\lambda}_{k,t}^{ID}}{\check{\lambda}_{k,t}^{ID}} y_{k,t}^{ID} \leq p_{k,t}^{ID} \Delta t \leq y_{k,t}^{ID} \qquad \forall k, t \geq \tau \quad (12b)$$

$$v_k^{ID}, \eta_{k,t}^{ID}, y_{k,t}^{ID} \geq 0 \qquad \forall k, t \geq \tau \quad (12c)$$

### 4.3. RVPP Deterministic Constraints

This section presents the deterministic constraints to formulate the RO problem for the RVPP market bidding. The supply-demand balancing constraints and the power traded constraints are presented in Sections 4.3.1 and 4.3.2, respectively.

#### 4.3.1. Supply-demand Balancing Constraints

The supply-demand balancing constraints for different market problems are presented in this section. Constraint (13) enforces the supply-demand balancing for the RVPP units connected to a single bus, considering both energy and up/down SR in the DAM+SRM. The reserve provision by each of the RVPP units is considered in the power balance constraint (13) by variables $r_{r,t}^{SR}$, $r_{\theta,t}^{SR}$, and $r_{d,t}^{SR}$. Three states are considered for variable $r_t^{SR}$, which is related to the total traded reserve by RVPP. When $r_t^{SR} = 0$, the supply-demand balancing constraints are only held for the power. When $r_t^{SR} = r_t^{SR,\uparrow}$ and $r_{r,t}^{SR} = -r_t^{SR,\downarrow}$, the supply-demand balancing constraints are held for the power and up SR reserve and power and down SR, respectively. Similar reserve activation scenarios are defined for units variables $r_{r,t}^{SR}$, $r_{\theta,t}^{SR}$, and $r_{d,t}^{SR}$ according to the above states. The above states keep the power balance equations for all possible SR situations in real-time (i.e., the SR not called on and up or/and down reserve called on). The goal is not to schedule the units according to the actual realization of the reserve, since it is rather difficult to accurately predict the sign and value of the SR activation. However, the logic behind these states is to assign power and SR boundaries (according to delivery time possibilities) to offer the best energy and reserve to the market.

$$\sum_{r \in R} \left( p_{r,t}^{DA} + r_{r,t}^{SR} \right) + \sum_{\theta \in \Theta} \left( p_{\theta,t}^{DA} + r_{\theta,t}^{SR} \right) = p_t^{DA} + r_t^{SR} + \sum_{d \in D} \left( p_{d,t}^{DA} - r_{d,t}^{SR} \right) \quad \forall t \quad (13)$$

The supply-demand balancing constraint in the DAM+SRM is fairly similar to SRM+IDM#1 and IDM#$k$ problems. If SRM+IDM#1 is solved, the power related to units in the DAM ($p_{r,t}^{DA*}$, $p_{\theta,t}^{DA*}$, and $p_{d,t}^{DA*}$) and the total power traded in the DAM ($p_{r,t}^{DA*}$) become parameters, and the variable related to the IDM#1 units power ($p_{(k=1),r,t}^{ID}$, $p_{(k=1),\theta,t}^{ID}$, and $p_{(k=1),d,t}^{ID}$) and total power traded in IDM#1 ($p_{(k=1),r,t}^{ID}$) are added to (14).

$$\sum_{r \in R} \left( p_{r,t}^{DA*} + p_{(k=1),r,t}^{ID} + r_{r,t}^{SR} \right) + \sum_{\theta \in \Theta} \left( p_{\theta,t}^{DA*} + p_{(k=1),\theta,t}^{ID} + r_{\theta,t}^{SR} \right) = p_t^{DA*} + p_{(k=1),t}^{ID} + r_t^{SR} + \sum_{d \in D} \left( p_{d,t}^{DA*} + p_{(k=1),d,t}^{ID} - r_{d,t}^{SR} \right) \quad \forall t \quad (14)$$

In IDM#$k$, in addition to power related to units in the DAM and the total power traded in the DAM, the reserve provided in the SRM+IDM#1 ($r_t^{SR*}$), the power related to units in the previous IDMs ($p_{k,r,t}^{ID*}$, $p_{k,\theta,t}^{ID*}$, and $p_{k,d,t}^{ID*}$), and the power traded in the previous IDMs ($p_{k,t}^{ID*}$) become parameters. Besides, the time periods for each session of the IDM#$k$ are updated by substituting (15) instead of (13).

$$\sum_{r \in R}\left(p_{r,t}^{DA*} + r_{r,t}^{SR} + \sum_{k=1}^{k-1} p_{k,r,t}^{ID*} + p_{k,r,t}^{ID}\right) + \sum_{\theta \in \Theta}\left(p_{\theta,t}^{DA*} + r_{\theta,t}^{SR} + \sum_{k=1}^{k-1} p_{k,\theta,t}^{ID*} + p_{k,\theta,t}^{ID}\right) =$$
$$p_t^{DA*} + r_t^{SR*} + \sum_{k=1}^{k-1} p_{k,t}^{ID*} + p_{k,t}^{ID} + \sum_{d \in D}\left(p_{d,t}^{DA*} - r_{d,t}^{SR} + \sum_{k=1}^{k-1} p_{k,d,t}^{ID*} + p_{k,d,t}^{ID}\right) \quad \forall k,t \geq \tau \quad (15)$$

### 4.3.2. Power Traded Constraints

The power traded constraints for different market snapshots are presented in this section. The equations (16a) and (16b) limit the maximum and minimum power and reserve to be traded in the DAM+SRM. The amount of requested down reserve by the TSO is a proportion of up reserve for each time period, modeled by (16c). Constraint (16d) sets the limit of the up reserve traded in the market according to a share of the total power production capacity of RVPP, defined by the user-defined parameter $\kappa$. Note that if $\kappa=0$, then the RVPP will only participate in the DAM, as no power would be allocated in the SRM.

$$p_t^{DA} + r_t^{SR,\uparrow} \leq \sum_{r \in R} \bar{P}_r + \sum_{\theta \in \Theta} \bar{P}_\theta \qquad \forall t \quad (16a)$$

$$-\sum_{d \in D} \bar{P}_d \leq p_t^{DA} - r_t^{SR,\downarrow} \qquad \forall t \quad (16b)$$

$$r_t^{SR,\uparrow} = \varrho_t r_t^{SR,\downarrow} \qquad \forall t \quad (16c)$$

$$r_t^{SR,\uparrow} \leq \kappa \left(\sum_{r \in R} \bar{P}_r + \sum_{\theta \in \Theta} \bar{P}_\theta\right) \qquad \forall t \quad (16d)$$

The power traded constraints in the SRM+IDM#1 are written in (17a)-(17c). The power that had already been traded in the DAM is fixed, and the IDM#1 power is added to these constraints.

$$p_t^{DA*} + p_{(k=1),t}^{ID} + r_t^{SR,\uparrow} \leq \sum_{r \in R} \bar{P}_r + \sum_{\theta \in \Theta} \bar{P}_\theta \qquad \forall t \quad (17a)$$

$$-\sum_{d \in D} \bar{P}_d \leq p_t^{DA*} + p_{(k=1),t}^{ID} - r_t^{SR,\downarrow} \qquad \forall t \quad (17b)$$

$$(16c)\text{-}(16d) \qquad (17c)$$

The power traded in IDM#k is calculated according to (18a)-(18b). The DAM power ($p_t^{DA*}$), the up and down reserve ($r_t^{SR,\uparrow*}$, $r_t^{SR,\downarrow*}$), and the power related to previous IDMs ($p_{k,t}^{ID*}$), which had already been assigned in the previous markets, are fixed in these equations.

$$p_t^{DA*} + \sum_{k=1}^{k-1} p_{k,t}^{ID*} + p_{k,t}^{ID} + r_t^{SR,\uparrow*} \leq \sum_{r \in R} \bar{P}_r + \sum_{\theta \in \Theta} \bar{P}_\theta \qquad \forall k,t \geq \tau \quad (18a)$$

$$-\sum_{d \in D} \bar{P}_d \leq p_t^{DA*} + \sum_{k=1}^{k-1} p_{k,t}^{ID*} + p_{k,t}^{ID} - r_t^{SR,\downarrow*} \qquad \forall k,t \geq \tau \quad (18b)$$

*4.4. RVPP Flexible Robust Constraints*

This section presents the robust constraints to formulate the RO problem for RVPP market bidding. With this aim, only the constraints that define the robustness of uncertain RVPP assets, such as ND-RESs, flexible demands, and STUs, are presented in Sections 4.4.1, 4.4.2, and 4.4.3, respectively. Deterministic operation constraints not modified by the RO (e.g., operation constraints of hydroelectric power plants, STUs, etc.) are omitted here. For the sake of conciseness, and without loss of generality, all RVPP units are assumed to be connected to a single bus, and network constraints are thus neglected. The robust constraints proposed in this section can be readily supplemented by network constraints in the form of well-known DC or AC power flow problems. Interested readers can find such constraints in, e.g., [2].

*4.4.1. ND-RESs Constraints*

The ND-RESs formulation in the DAM and SRM is presented in (19a)-(19k). Constraints (19a) and (19b) limit the up/down reserve that each ND-RES can provide by considering a ramp-constraint response of the ND-RES and the required time for SR action presented by the TSO. The lower limit of ND-RESs output power is given by (19c). Constraint (19d) is the upper limit of ND-RESs production by considering the median and negative deviation of the power forecast. Only negative deviations are considered in the formulation to define worst-case scenarios, as positive deviations will always benefit the RVPP. Constraints (19e)-(19k) apply the robust formulation to the uncertain parameters for stochastic ND-RESs generation similar to (5a)-(5h) in Section 3.2. The binary variable $\chi_{r,t}^{DA}$ sets the active or non-active status of each period to satisfy the predefined robustness budget. In this work, $\Gamma_r^{DA}$ is included in the formulation to make the highest energy reduction scenario in the whole operation horizon flexible, as a robustness budget for a global scheduling horizon. Constraint (19e) assigns the maximum value for power reduction in each period according to the uncertainty budget $\Gamma_r^{DA}$ that the RVPP operator predefines for the entire operation horizon. The dual variables related to ND-RESs energy uncertainties in (19f), $v_r^{DA}$ and $\eta_{r,t}^{DA}$, set a lower bound on energy deviation. Both dual variables are bounded by (19g) and logically constrained by (19h) based on the active or non-active status of the periods to comply with the robustness budget. If the robust status of a period is active ($\chi_{r,t}^{DA} = 1$), constraint (19h) leads to a reduction of input energy of ND-RESs on that specific period by allowing a positive amount for $\eta_{r,t}^{DA}$. Constraint (19i) limits the number of periods affected by considering the robust strategy. For instance, if $\Gamma_r^{DA} = 3$, the ND-RES power deviation happens in 3 hours that results in the worst energy reduction. Finally, the nature of positive auxiliary variables is defined in (19j), whereas constraint (19k) shows the nature of auxiliary binary variables.

$$r_{r,t}^{SR,\uparrow} \leq T^{SR}\overline{R}_r^{SR} \qquad \forall r,t \quad (19a)$$

$$r_{r,t}^{SR,\downarrow} \leq T^{SR}\underline{R}_r^{SR} \qquad \forall r,t \quad (19b)$$

$$\underline{P}_r \leq p_{r,t}^{DA} - r_{r,t}^{SR,\downarrow} \qquad \forall r,t \quad (19c)$$

$$p_{r,t}^{DA} + r_{r,t}^{SR,\uparrow} \leq \widetilde{P}_{r,t}^{DA} - y_{r,t}^{DA} \qquad \forall r,t \quad (19d)$$

$$y_{r,t}^{DA} \leq \breve{P}_{r,t}^{DA} \qquad \forall r,t \quad (19e)$$

$$y_{r,t}^{DA} \geq v_r^{DA} + \eta_{r,t}^{DA} - M\left(1 - \chi_{r,t}^{DA}\right) \qquad \forall r,t \quad (19f)$$

$$v_r^{DA} + \eta_{r,t}^{DA} \geq \breve{P}_{r,t}^{DA} \qquad \forall r,t \quad (19g)$$

$$\varepsilon \chi_{r,t}^{DA} \leq \eta_{r,t}^{DA} \leq M \chi_{r,t}^{DA} \qquad \forall r,t \quad (19h)$$

$$\sum_t \chi_{r,t}^{DA} = \Gamma_r^{DA} \qquad \forall r \quad (19i)$$

$$v_r^{DA}, \eta_{r,t}^{DA}, y_{r,t}^{DA} \geq 0 \qquad \forall r,t \quad (19j)$$

$$\chi_{r,t}^{DA} \in \{0,1\} \qquad \forall r,t \quad (19k)$$

The robust formulation of ND-RESs in the SRM+IDM#1 and IDM#$k$ problems is fairly similar to the DAM+SRM. If SRM+IDM#1 is solved, the power traded in the DAM ($p_{r,t}^{DA*}$) becomes a parameter, and the variable related to the IDM#1 power ($p_{(k=1),r,t}^{ID}$) is added to (19c) and (19d). Besides, the parameters related to the median/negative deviation of the ND-RES $r$ production forecast in the SRM ($\widetilde{P}_{r,t}^{SR}/\breve{P}_{r,t}^{SR}$) are updated for SRM+IDM#1 by substituting (20b), (20c), and (20d) instead of (19d), (19e), and (19g), respectively.

$$\underline{P}_r \leq p_{r,t}^{DA*} - r_{r,t}^{SR,\downarrow} + p_{(k=1),r,t}^{ID} \qquad \forall r,t \quad (20a)$$

$$p_{r,t}^{DA*} + r_{r,t}^{SR,\uparrow} + p_{(k=1),r,t}^{ID} \leq \widetilde{P}_{r,t}^{SR} - y_{r,t}^{SR} \qquad \forall r,t \quad (20b)$$

$$y_{r,t}^{SR} \leq \breve{P}_{r,t}^{SR} \qquad \forall r,t \quad (20c)$$

$$v_r^{SR} + \eta_{r,t}^{SR} \geq \breve{P}_{r,t}^{SR} \qquad \forall r,t \quad (20d)$$

In IDM#$k$, in addition to the power traded in the DAM, the power in the previous IDM#1-IDM#$k$-1 ($p_{k,r,t}^{ID*}$) becomes a parameter. The forecast parameters of ND-RES $r$ production in the IDM ($\widetilde{P}_{k,r,t}^{ID}/\breve{P}_{k,r,t}^{ID}$) are updated. Besides, the time periods for each session of the IDM#$k$ are updated by substituting (21a)-(21d) instead of (19c), (19d), (19e), and (19g).

$$\underline{P}_r \leq p_{r,t}^{DA*} - r_{r,t}^{SR,\downarrow} + \sum_{k=1}^{k-1} p_{k,r,t}^{ID*} + p_{k,r,t}^{ID} \qquad \forall k,r,t \geq \tau \quad (21a)$$

$$p_{r,t}^{DA*} + r_{r,t}^{SR,\uparrow} + \sum_{k=1}^{k-1} p_{k,r,t}^{ID*} + p_{k,r,t}^{ID} \leq \widetilde{P}_{k,r,t}^{ID} - y_{k,r,t}^{ID} \qquad \forall k,r,t \geq \tau \quad (21b)$$

$$y_{k,r,t}^{ID} \leq \breve{P}_{k,r,t}^{ID} \qquad \forall k,r,t \geq \tau \quad (21c)$$

$$v_{k,r}^{ID} + \eta_{k,r,t}^{ID} \geq \breve{P}_{k,r,t}^{SR} \qquad \forall k,r,t \geq \tau \quad (21d)$$

### 4.4.2. Flexible Demands Constraints

The robust constraints related to the flexible demands are formulated in this section. This formulation is an upgrade of the deterministic model of flexible demands in VPPs presented in [37]. There, the authors proposed a bi-level flexibility model for demands based on selecting a consumption profile from a pre-defined set in the DAM (first level) and the definition of thresholds around the selected profile for varying the consumption in SRM and IDMs (second level).

Constraint (22a) assigns the demand for each period to predefined profiles of demands considering the median and positive deviation of demand. Constraint (22b) assures that only one profile between several profiles of the load is selected by the algorithm. The maximum positive deviation of demand is limited by (22c) and (22d). Constraints (22a)-(22d) are elaborated based on the approach of selecting the predefined profiles of demands in [37] and by applying the robust method analogous to constraints (6a) and (6b) in Section 3.2. Constraints (22e)-(22h) are defined similarly to robust constraints (5c), (6c), (5e), and (5f) in Section 3.2 to assign the maximum value for demand in each period according to the uncertainty budget $\Gamma_d^{DA}$ which RVPP operator defines. Constraints (22i)-(22l) limit the up/down reserve by a specified amount of load (flexibility of demand) selected by constraint (22a), considering that the uncertainty can affect demand in some periods. The ramp-up and ramp-down limitations of load are restricted by (22m) and (22n), respectively. These constraints define the worst condition of the ramp rate of the load when the up or down reserve is activated in two sequential periods. Constraints (22o) and (22p) bound the load capability for providing the up/down reserve. Constraint (22q) confines the minimum energy that the demand should use. This equation is written for the worst condition of providing reserve from the minimum daily energy consumption perspective, i.e., when the demand provides the up reserve which requires the demand to further reduce its consumption. Finally, constraints (22r) and (22s) define the nature of positive and binary variables, respectively. The robust formulation of flexible demands in the SRM+IDM#1 and IDM#k can be deduced following the same process as for ND-RESs in Section 4.4.1, and their equations are omitted here.

$$p_{d,t}^{DA} = \sum_{p \in P} \left( \widetilde{P}_{d,p,t}^{DA} u_{d,p} \right) + y_{d,t}^{DA} \qquad \forall d,t \qquad (22a)$$

$$\sum_{p \in P} u_{d,p} = 1 \qquad \forall d \qquad (22b)$$

$$y_{d,t}^{DA} \leq \sum_{p \in P} \widehat{P}_{d,p,t}^{DA} u_{d,p} \qquad \forall d,t \qquad (22c)$$

$$y_{d,t}^{DA} \leq M \chi_{d,t}^{DA} \qquad \forall d,t \qquad (22d)$$

$$y_{d,t}^{DA} \geq v_d^{DA} + \eta_{d,t}^{DA} - M\left(1 - \chi_{d,t}^{DA}\right) \qquad \forall d,t \qquad (22e)$$

$$v_d^{DA} + \eta_{d,t}^{DA} \geq \sum_{p \in P} \widehat{P}_{d,p,t}^{DA} u_{d,p} \qquad \forall d,t \qquad (22f)$$

$$\varepsilon \chi_{d,t}^{DA} \leq \eta_{d,t}^{DA} \leq M \chi_{d,t}^{DA} \qquad \forall d,t \qquad (22g)$$

$$\sum_t \chi_{d,t}^{DA} = \Gamma_d^{DA} \qquad \forall d \qquad (22h)$$

$$r_{d,t}^{SR,\uparrow} \leq \underline{\beta}_{d,t} \left( p_{d,t}^{DA} - y_{d,t}^{DA} \right) \qquad \forall d,t \qquad (22i)$$

$$r_{d,t}^{SR,\uparrow} \leq p_{d,t}^{DA} - \underline{P}_d \qquad \forall d,t \qquad (22j)$$

$$r_{d,t}^{SR,\downarrow} \leq \overline{\beta}_{d,t} \left( p_{d,t}^{DA} - y_{d,t}^{DA} \right) \qquad \forall d,t \qquad (22k)$$

$$r_{d,t}^{SR,\downarrow} \leq \overline{P}_d - p_{d,t}^{DA} \qquad \forall d,t \quad (22l)$$

$$\left(p_{d,t}^{DA} + r_{d,t}^{SR,\downarrow}\right) - \left(p_{d,(t-1)}^{DA} - r_{d,(t-1)}^{SR,\uparrow}\right) \leq \overline{R}_d \Delta t \qquad \forall d,t \quad (22m)$$

$$\left(p_{d,(t-1)}^{DA} + r_{d,(t-1)}^{SR,\downarrow}\right) - \left(p_{d,t}^{DA} - r_{d,t}^{SR,\uparrow}\right) \leq \underline{R}_d \Delta t \qquad \forall d,t \quad (22n)$$

$$r_{d,t}^{SR,\uparrow} \leq T^{SR} \underline{R}_d^{SR} \qquad \forall d,t \quad (22o)$$

$$r_{d,t}^{SR,\downarrow} \leq T^{SR} \overline{R}_d^{SR} \qquad \forall d,t \quad (22p)$$

$$\underline{E}_d \leq \sum_{t \in T} \left(p_{d,t}^{DA} \Delta t - r_{d,t}^{SR,\uparrow}\right) \qquad \forall d \quad (22q)$$

$$v_d^{DA}, \eta_{d,t}^{DA}, y_{d,t}^{DA} \geq 0 \qquad \forall d,t \quad (22r)$$

$$\chi_{d,t}^{DA} \in \{0,1\} \qquad \forall d,t \quad (22s)$$

*4.4.3. Solar Thermal Units Constraints*

STUs are a particular type of ND-RES, as they also include molten-salt energy storage capability, and thus, they cannot be accurately represented by the model in 4.4.1. In STUs, the source of uncertainty, i.e., solar irradiation, affects the thermal output power at the solar field, which is then used to either convert it to electric power to be injected into the grid or used to charge the molten-salt storage device. The efficiency of the conversion between thermal and electric powers depends on the thermal power passed through the Power Block of the STU, with typical values of 30-40% (see [38] for more details). The full deterministic model of the STU with its storage can be found in [2] and [35]. As justified at the beginning of Section 4.4, only those constraints in the model affected by the robust formulations are presented in (23a)-(23h). Again, the robust formulations of STUs in the SRM+IDM#1 and IDM#$k$ are fairly similar to the DAM+SRM problem, and their equations are omitted.

$$0 \leq p_{\theta,t}^{SF} \leq \widetilde{P}_{\theta,t}^{DA} - y_{\theta,t}^{DA} \qquad \forall \theta,t \quad (23a)$$

$$y_{\theta,t}^{DA} \leq \breve{P}_{\theta,t}^{DA} \qquad \forall \theta,t \quad (23b)$$

$$y_{\theta,t}^{DA} \geq v_\theta^{DA} + \eta_{\theta,t}^{DA} - M\left(1 - \chi_{\theta,t}^{DA}\right) \qquad \forall \theta,t \quad (23c)$$

$$v_\theta^{DA} + \eta_{\theta,t}^{DA} \geq \breve{P}_{\theta,t}^{DA} \qquad \forall \theta,t \quad (23d)$$

$$\varepsilon \chi_{\theta,t}^{DA} \leq \eta_{\theta,t}^{DA} \leq M\chi_{\theta,t}^{DA} \qquad \forall \theta,t \quad (23e)$$

$$\sum_t \chi_{\theta,t}^{DA} = \Gamma_\theta^{DA} \qquad \forall \theta \quad (23f)$$

$$v_\theta^{DA}, \eta_{\theta,t}^{DA}, y_{\theta,t}^{DA} \geq 0 \qquad \forall \theta,t \quad (23g)$$

$$\chi_{\theta,t}^{DA} \in \{0,1\} \qquad \forall \theta,t \quad (23h)$$

After thoroughly describing the robust RVPP bidding optimization algorithm in (1)-(23) as a manageable, single-level MILP problem, several case studies are then explored in Section 5. These case studies have been designed to test and validate the algorithm's efficacy and practical applicability.

## 5. Case Studies

In this section, a set of case studies is presented to evaluate the proposed flexible RO model for RVPP market bidding. To this aim, different RVPP configurations and electricity market sequences are considered. The RVPP may include a wind farm, a solar PV plant, an STU, and/or flexible demand. A residential aggregator is considered for the flexible demand, with three different profiles available for choice at the DAM, as in [37]. The minimum daily consumption of each profile is 360 MWh. The maximum possible demand value at each period is 30 MW. During SRM and IDMs, the demand owner allows a $\pm 10\%$ tolerance for additional demand flexibility over the selected profile at DAM.

The forecast data of the wind farm and solar PV energy production available to solve the DAM are depicted in Figure 3. Only negative deviations are evaluated, as discussed in Section 4.4. Updates of such uncertain profiles for subsequent market sessions, as well as for the STU, are defined similarly and not shown here. The solar PV and the wind farm both have a rated capacity of 50 MW each, and their operation costs are respectively 5 €/MWh and 10 €/MWh. The STU power block has a rated electrical capacity of 50 MW, and the thermal storage capacity of the STU is 1100 MWh. The operation cost of the STU is 15 €/MWh. The price forecast data for DAM, SRM, and IDM#1 are adopted from the REE website, and the former is shown in Figure 4 for the sake of illustration [39], where the asymmetric distribution is apparent.

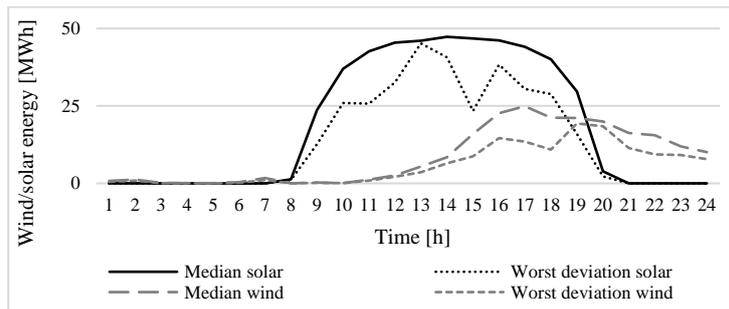

Figure 3 Median/worst case of wind/solar production in the DAM.

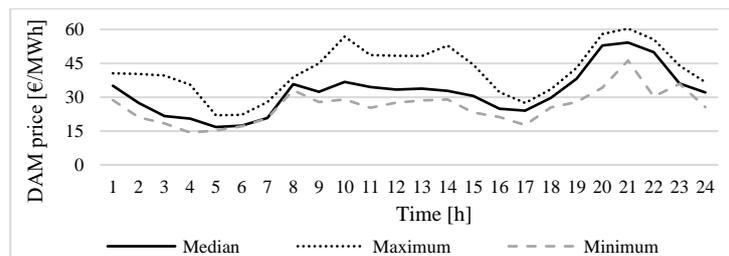

Figure 4 DAM electricity price forecast range [39].

Table 1 shows the RVPP configuration and uncertainty data included in each case study. In *Case 1*, the capability of the proposed model to handle the asymmetric uncertainties in the electricity prices and to consider the whole period for energy robustness compared to the robust formulation proposed in [23] is evaluated. In *Case 2*, multiple uncertainties in the ND-RESs production, flexible demand, and electricity market prices (DAM & SRM) are considered to show the effectiveness of the proposed model. Besides, the sensitivity analysis for different uncertainty budgets is done to evaluate the behavior of RVPP operator considering risk. In *Case 3*, the bidding strategy of RVPP in the sequential markets is analyzed by considering the updated forecast data in the SRM and IDMs.

The simulations are carried out using a Dell XPS with an i7-1165G7 processor, 2.8 GHz, and 16 GB of RAM using the CPLEX solver in GAMS 38.3.0. The simulation time for all case studies is below 2 s.

| Case | RVPP configuration | | | | Uncertain data | | | | | | |
|---|---|---|---|---|---|---|---|---|---|---|---|
| | Wind | Solar | STU | Demand | Wind | Solar | STU | Demand | DAM | SRM | IDMs |
| 1 | ✓ | ✓ | ✓ | ✓ | ✓ | ✓ | ✓ | | ✓ | ✓ | |
| 2 | ✓ | ✓ | ✓ | ✓ | ✓ | ✓ | ✓ | ✓ | ✓ | ✓ | |
| 3 | ✓ | ✓ | | ✓ | ✓ | ✓ | | | ✓ | ✓ | ✓ |

Table 1 RVPP configuration and uncertainty data of each case study.

### 5.1. Case 1

In the first analysis of *Case 1*, only uncertainties in electricity prices are considered. Figure 5 compares the DAM traded power for the proposed model and the model in [23] by considering uncertainties in the DAM and SRM price ($\Gamma^{DA} = 5$ and $\Gamma^{SR,\uparrow}/\Gamma^{SR,\downarrow} = 5$). The figure also depicts the maximum and minimum deviation of the DAM price forecast compared to the median and mean values. Figure 6 presents and compares the traded up/down reserve, as well as the SRM price and its deviation. In [23], the asymmetry in the electricity prices is not considered, so instead of the median value of forecast data, the mean value is used in the objective function of the optimization problem. Considering different electricity prices leads to some differences between the results of both models. The proposed model results in more conservative results in hours 10, 12, 13, and 14; therefore, the RVPP sells less energy in the electricity market. For instance, the traded energy in hours 10 and 14 is 22% and 40% lower than the model proposed in [23]. However, the proposed model results in selling more energy in the DAM in hours 11, 16, 17, and 22. In hour 17, the sold energy in the model proposed in [23] is 40% lower than the proposed model. These differences between traded energy in both models are due to the variation of electricity prices that in turn comes from uncertainty happening in different hours. Therefore, considering lower electricity prices in hour 17 in the model proposed in [23] due to uncertainty (the minimum electricity price), the STU in the RVPP stores energy in this hour to provide more up reserve in hour 22 according to Figure 6.

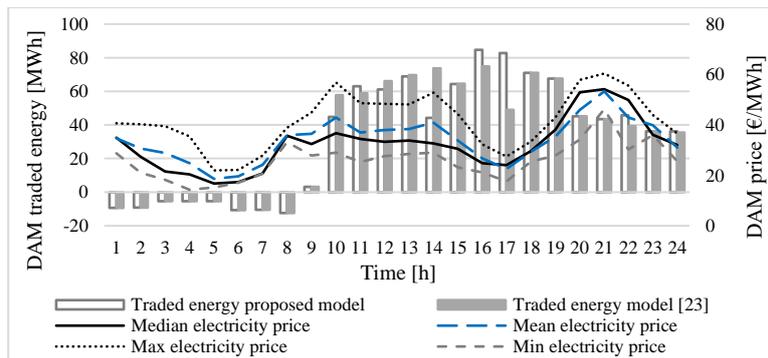

Figure 5 DAM electricity price forecast range and traded power in *Case 1* for the proposed model and the model in [23] ($\Gamma^{DA} = 5$ and $\Gamma^{SR,\uparrow}/\Gamma^{SR,\downarrow} = 5$).

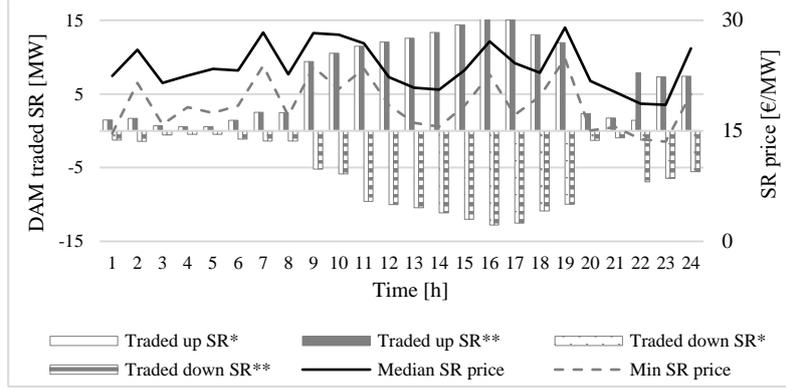

Figure 6 SR price forecast range and traded SR in *Case 1* for the proposed model (*) and the model in [23] (**) in DAM ($\varGamma^{DA} = 5$ and $\varGamma^{SR,\uparrow}/\varGamma^{SR,\downarrow} = 5$).

Figure 7 compares the DAM profit in the proposed model and the model in [23] for different combinations and values of uncertainty budgets. These include the *energy robustness case* (wind and solar production), the DAM *price robustness case*, and the *price-energy robustness case*. The energy uncertainty budget in the proposed model ($\varGamma^{DA}_{r(\theta)}$) and the model in [23] ($\varGamma^{DA}_{r(\theta),t}$) changes between 0-24 and 0-1, respectively. Therefore, the uncertainty budget is shown in relative terms with respect to the maximum budget. Price uncertainty budgets for both models ($\varGamma^{DA}$) are similar and change between 0-24. In both models, increasing the uncertainty budget decreases the DAM profit for the above three cases. However, in the proposed model, the saturation of RVPP profit happens at smaller percentages of uncertainty budget, i.e., worst realizations are captured first. For example, for an uncertainty budget of 21% for the *price robustness case*, the RVPP profit compared to the deterministic case is reduced by 19% and 13% in the proposed model and the model [23], respectively. Besides, the figure shows that in the model [23], the *energy robustness case* is linear, and *price* and *price-energy cases* are almost linear. This is due to the fact that by increasing the energy uncertainty budget, which has a significant effect compared to the price uncertainty budget on RVPP profit, the energy is reduced evenly in all time periods.

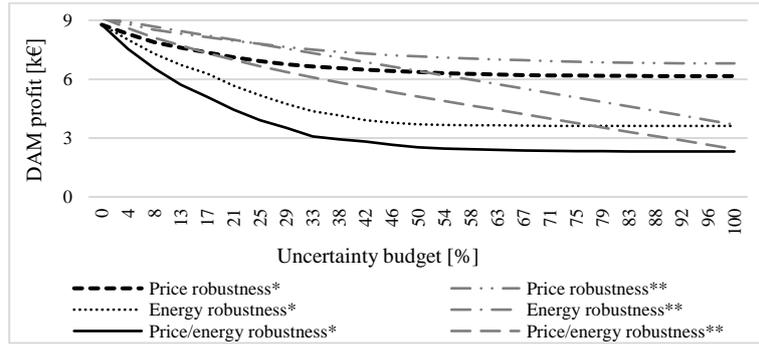

Figure 7 DAM profit in *Case 1* for the proposed model (*) and the model in [23] (**) for various uncertainty budgets:
    **Energy robustness:** ($\varGamma^{DA}_{r(\theta)} \in$ [0-24] or $\varGamma^{DA}_{r(\theta),t} \in$ [0-1], $\varGamma^{DA} = 0$);
    **Price robustness:** ($\varGamma^{DA}_{r(\theta)} = 0$ or $\varGamma^{DA}_{r(\theta),t} = 0$, $\varGamma^{DA} \in$ [0-24]);
    **Price-energy robustness:** ($\varGamma^{DA}_{r(\theta)} = $ [0-24] or $\varGamma^{DA}_{r(\theta),t} \in$ [0-1], $\varGamma^{DA} \in$ [0-24]).

In the second analysis of *Case 1*, only uncertainties in units' energy are considered. Figure 8 compares the possible unfeasible region of wind farm available energy in the deterministic (optimistic) case, in the proposed model for $\varGamma^{DA}_r = 5$, and in the model [23] for $\varGamma^{DA}_{r,t} = 5/24$. The possible unfeasible region corresponds to the area between the traded energy and the minimum available energy. These unfeasible regions are depicted using different backgrounds in the figure. By assuming an accurate forecast bound for wind production, unfeasibility can occur if the wind production is less than the median wind production minus the deviation value. In the deterministic case, a relatively large possible unfeasible region is observable according to Figure 8 since any negative deviation during the scheduling period can lead to unfeasibility or penalty for RVPP. In the proposed model, the possible unfeasible region is significantly reduced by capturing the worst case of wind deviation instead of considering even fluctuation.

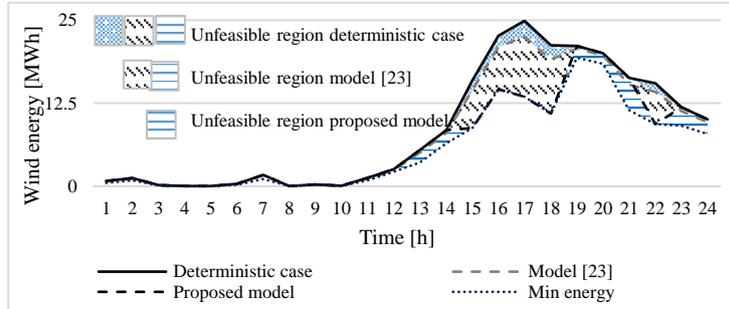

Figure 8 Possible unfeasible region of wind farm energy in *Case 1* in the deterministic model, proposed model ($\Gamma_r^{DA} = 5$) and the model in [23] ($\Gamma_{r,t}^{DA} = 5/24$).

Finally, in the last analysis of *Case 1*, the uncertainties in both energy and electricity prices are considered. An out-of-sample assessment is performed (see Appendix C) to further justify the proposed robust model compared to the model in [23]. One hundred scenarios are generated based on the hourly distributions of different uncertain parameters for all cases discussed below. To better capture the asymmetric behavior of different distributions, the Weibull distribution is used to generate the scenarios. The Weibull distribution is a versatile probability distribution capable of modeling various data with varying degrees of skewness and tail behavior. These characteristics make it a valuable tool in statistical analyses. The penalty parameter *Z* is set to 1000 €/MWh as used in [26] by the same authors of the approach proposed in [23], to compare better the results of such an approach with the model proposed in this paper. The results of out-of-sample assessment for different uncertainty budgets are provided in Table 2. The results compare the average operating profit of RVPP (it does not include the penalization cost) in all scenarios defined by $\Pi^{av}$, the average penalization cost $K^{av}$, and the net profit $\Pi^{av} - K^{av}$ of the RVPP. Due to the limited number of hours with renewable production and the low probability of having more than 6 simultaneous hours out of 24 hours at the worst deviation, the table only presents the results for the uncertainty budget between 0-6.

The results show that the average penalization cost is very high for the deterministic approach (when all uncertainty budgets are zero). This leads to a significant loss for RVPP according to the first row of Table 2. The model [23] obtains more average operating profit at higher uncertainty budget values than the proposed model. This is due to the fact that the proposed approach is more conservative in capturing the worst-case scenarios. Besides, increasing the uncertainty budget reduces the value of the average penalization cost in both the proposed model and the model [23]. However, considering the worst case of energy uncertainty, the proposed model has a better performance in terms of net profit and average penalization cost. It is worth noting that higher penalizations on some electric markets, such as the SRM, could lead the RVPP to be excluded from participation in those markets. The proposed approach results in higher net profit for most uncertainty budgets (1-5). The net profit in the proposed model is 57.5%, 89.3%, 90.6%, 69.8%, and 10.7% higher than the net profit in model [23]. Only for uncertainty budget 6, the net profit of model [23] is 75.7% higher than the net profit of proposed model. This is due to the fact that in the higher uncertainty budgets, the average penalization cost is a smaller proportion of the net profit of RVPP. Note that the penalization cost in the proposed model for uncertainty budgets 3 and 4 are equal. This is due to the flexibility of thermal storage of STU that can reduce it in most of the hours that STU is producing energy.

| Uncertainty budget | Proposed model | | | Model [23] | | |
|---|---|---|---|---|---|---|
| | $\Pi^{av}$ [€] | $K^{av}$ [€] | $\Pi^{av} - K^{av}$ [€] | $\Pi^{av}$ [€] | $K^{av}$ [€] | $\Pi^{av} - K^{av}$ [€] |
| 0 | 24947 | 140762 | -115815 | 24874 | 140802 | -115928 |
| 1 | 22364 | 61383 | -39019 | 24090 | 115902 | -91812 |
| 2 | 20315 | 27533 | -7218 | 23300 | 91281 | -67981 |
| 3 | 19026 | 23205 | -4179 | 22437 | 67049 | -44612 |
| 4 | 17506 | 23205 | -5699 | 21712 | 40598 | -18886 |
| 5 | 16034 | 20365 | -4331 | 20858 | 25708 | -4850 |
| 6 | 14493 | 15833 | -1340 | 20336 | 20661 | -325 |

Table 2 The out-of-sample assessment for proposed model and the model in [23] ($\Gamma^{DA} = \Gamma^{SR,\uparrow} = \Gamma^{SR,\downarrow} = [0\text{-}6]$, $\Gamma^{DA}_{r(\theta)} = [0\text{-}6]$ or $\Gamma^{DA}_{r(\theta),t} = [0\text{-}0.25]$).

To further verify the results obtained in Table 2, the frequency histograms of defined variables for each scenario and uncertainty budget 2 are depicted in Figure 9, Figure 10, and Figure 11. Figure 9 shows the distribution of operating profit of RVPP without considering the penalization cost. Figure 10 and Figure 11 show the penalization cost and net profit, respectively. The operating profit in the model [23] is more than the proposed model in most scenarios according to Figure 9. However, according to Figure 10 and Figure 11, the proposed model has better performance when the penalization cost is considered.

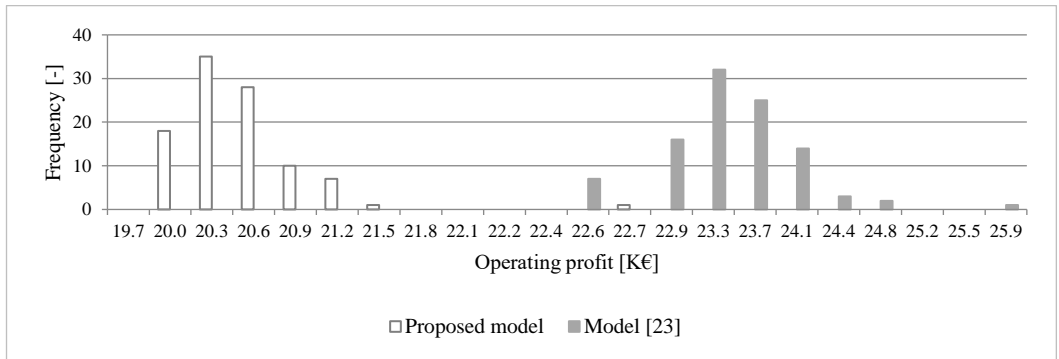

Figure 9 Frequency histogram of sampled daily profit for the proposed model and the model in [23] ($\Gamma^{DA} = \Gamma^{SR,\uparrow} = \Gamma^{SR,\downarrow} = 2$, $\Gamma^{DA}_{r(\theta)} = 2$ or $\Gamma^{DA}_{r(\theta),t} = 2/24$).

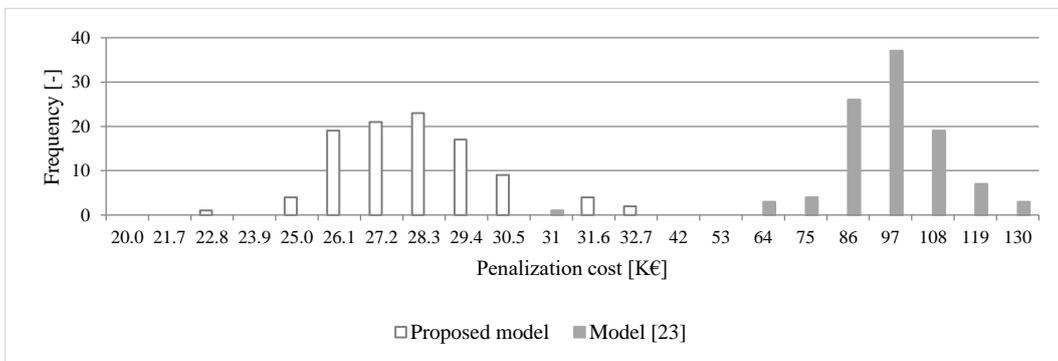

Figure 10 Frequency histogram of sampled daily imbalance cost for the proposed model and the model in [23] ($\Gamma^{DA} = \Gamma^{SR,\uparrow} = \Gamma^{SR,\downarrow} = 2$, $\Gamma^{DA}_{r(\theta)} = 2$ or $\Gamma^{DA}_{r(\theta),t} = 2/24$).

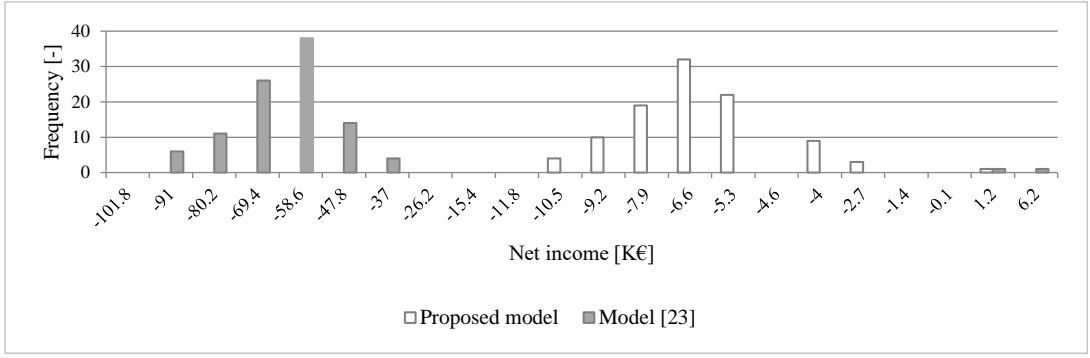

Figure 11 Frequency histogram of sampled daily final profit for the proposed model and the model in [23] ($\Gamma^{DA} = \Gamma^{SR,\uparrow} = \Gamma^{SR,\downarrow} = 2$, $\Gamma^{DA}_{r(\theta)} = 2$ or $\Gamma^{DA}_{r(\theta),t} = 2/24$).

## 5.2. Case 2

Figure 12 shows the DAM profit when only energy uncertainties (wind and solar production, STU thermal production, and demand) are considered, when only price uncertainties (DAM and SRM) are considered, and when all uncertainties, both in price and energy, are included. The RVPP income reduction at the saturation point, i.e., when all uncertainty budgets are set to 10 compared to total possible income reduction (uncertain parameters=24) for the three cases, *energy-robustness*, *price robustness*, and *price-energy robustness*, is 92%, 73%, and 90%, respectively. The larger percentage of income reduction for the *energy-robustness* case is due to the fact that the forecasted production of ND-RES and STU units is zero or almost zero in some hours. Therefore, the saturation of income reduction happens at lower values of uncertainty budget compared to the *price robustness* case. In the *price robustness case*, the DAM and SRM price uncertainty budgets are increased from 0 to 24 in unity steps. By increasing uncertainty budgets, the RVPP income decreases with a higher slope for lower values of uncertainty budgets than for greater ones. This trend is expected as the worst cases of price deviations are selected first in the robust formulation. Comparing all cases, it can be observed that, in the scenario considered, the majority of the RVPP profit reduction comes from energy uncertainty. Indeed, all RVPP units have uncertain productions, and reducing them has a higher impact on the RVPP income than price fluctuation.

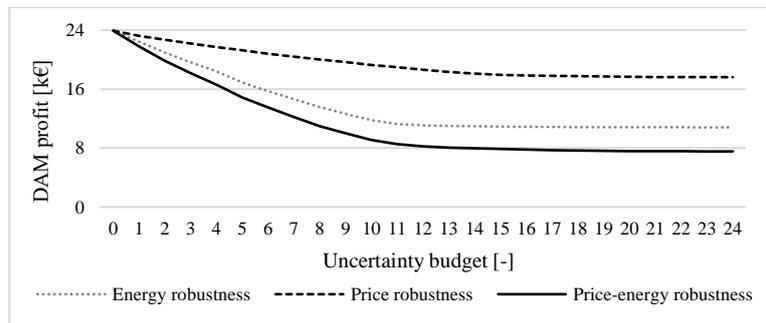

Figure 12 DAM profit in *Case 2* for various uncertainty budgets:
  **Energy robustness:** ($\Gamma^{DA}_{r(\theta)} = \Gamma^{DA}_d \in [0\text{-}24]$, $\Gamma^{DA} = \Gamma^{SR,\uparrow} = \Gamma^{SR,\downarrow} = 0$);
  **Price robustness:** ($\Gamma^{DA}_{r(\theta)} = \Gamma^{DA}_d = 0$, $\Gamma^{DA} = \Gamma^{SR,\uparrow} = \Gamma^{SR,\downarrow} \in [0\text{-}24]$);
  **Price-energy robustness:** ($\Gamma^{DA}_{r(\theta)} = \Gamma^{DA}_d = \Gamma^{DA} = \Gamma^{SR,\uparrow} = \Gamma^{SR,\downarrow} \in [0\text{-}24]$).

The analysis at the RVPP unit level is next discussed. To this aim, Figure 13 shows DAM traded energy, the wind farm, STU, and solar energy for a deterministic case. Figure 14 depicts the same results as Figure 13 when the uncertainty budgets for energy (wind farm, STU, and solar energy) and price (DAM and SRM) are set to 5. The total sold/bought energy in the uncertain case is decreased/increased by 41% and 8% compared to the deterministic case. By increasing the uncertainty budget, the hours that impose the highest deviations in energy according to the energy forecast are selected as the worst cases. For instance, hours 15-18 and 22 for wind production, hours 11, 12, 15, 17, and 19 for available solar energy, hours 10-14 and 15 for STU available energy, and hours 8, 15-17, and 19 for demand are the worst hours from a deviation perspective for each technology.

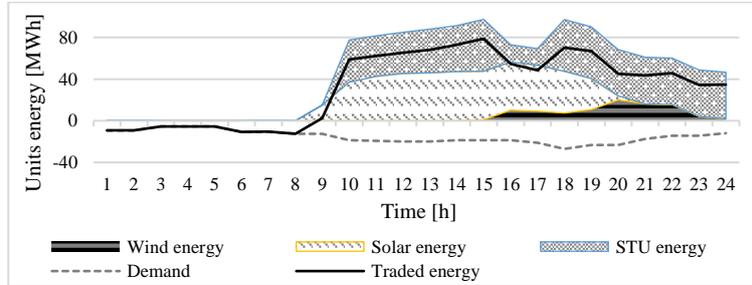

Figure 13 RVPP different units energy and traded energy in *Case 2* in the DAM for deterministic case.

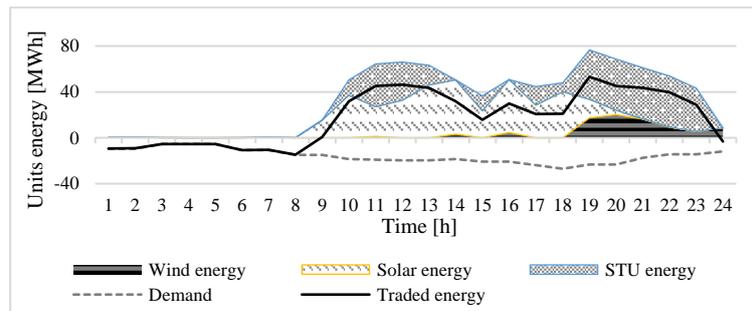

Figure 14 RVPP different units energy and traded energy in *Case 2* in the DAM for $\Gamma_{r(\theta)}^{DA} = \Gamma^{DA} = \Gamma^{SR,\uparrow} = \Gamma^{SR,\downarrow} = 5$.

### 5.3. Case 3

Finally, the flexible robust optimal participation of an RVPP in the complete sequence of market sessions is considered. This case study assumes enough liquidity on IDMs to show the RVPP model capabilities. For instance, the first IDM (IDM#1) in the Spanish market has high liquidity in some circumstances. Figure 15 displays the total traded energy and SR for the deterministic case by assuming RVPP participates in DAM, SRM, IDM#1, and IDM#4. In the deterministic case, RVPP sells most of its available energy in the DAM and IDM#1 between hours 9-22. Besides, the RVPP mostly buys energy in the DAM in hours 1-8 and 23-24. However, the RVPP sells energy in hours 23-24 in IDM#1 to arbitrage energy between DAM and IDM. There is also a small percentage of traded energy in IDM#4 compared to the previous markets, as a small value of energy is usually traded in the last sessions of IDMs. The RVPP trades both up and down SR in the SRM in hours 9-24 as it has available wind or/and solar production.

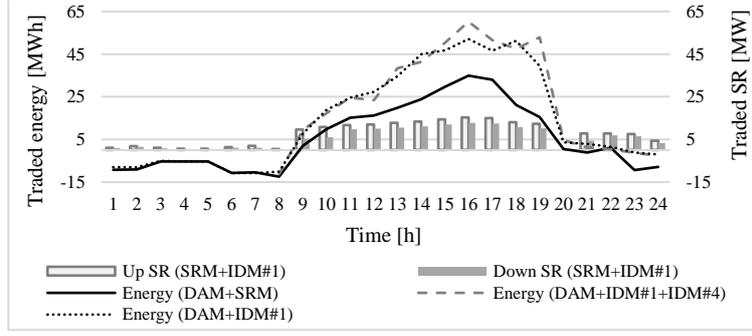

Figure 15 Total traded energy and SR in *Case 3* for deterministic case.

Figure 16 shows the same results as Figure 15 when the uncertainty budgets for energy (wind and solar) and all price uncertainty budgets are equal to 5. In this case, the RVPP prefers to sell less energy in the DAM due to the uncertainty. In the uncertain case, the total sold/bought energy until the last considered market session (IDM#4) is decreased/increased by 28% and 14% compared to the deterministic case. In hour 15, it sells almost nothing in the DAM and mostly provides the up and down reserve. However, when it goes to IDM#1 with a more precise forecast value, the traded energy in hour 15 is increased. This is also due to the fact that in hour 15, the DAM price is low; however, the IDM price is relatively high, so RVPP has an opportunity to adjust its traded energy. The same trend is understandable for traded energy in IDM#1 compared to DAM between hours 9-19. The amount of trading up and down SR is also decreased compared to the deterministic one as a more conservative strategy is adopted.

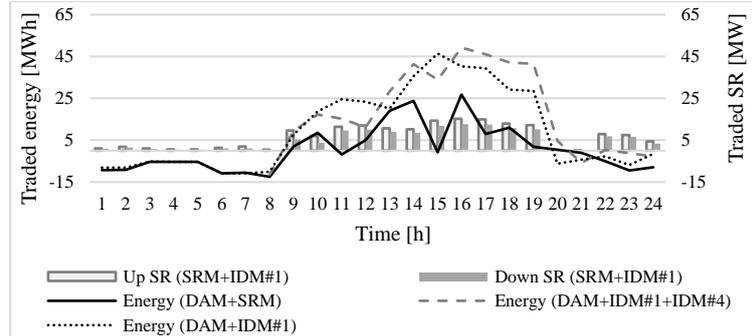

Figure 16 Total traded energy and SR in *Case 3* for $\varGamma_r^{DA} = \varGamma^{DA} = \varGamma^{SR,\uparrow} = \varGamma^{SR,\downarrow} = \varGamma_{(k=1)}^{ID} = \varGamma_{(k=4)}^{ID} = 5$.

## 6. Conclusion

This paper proposes a novel mathematical formulation of the optimization problem for renewable-only RVPP participation in sequential energy and reserve markets. A new and efficient flexible robust optimization approach is implemented as an MILP to accurately capture the asymmetric uncertainties in the electricity prices and RESs in the day-ahead, secondary reserve, and intra-day markets. Moreover, as opposed to other robust optimization models, the proposed approach considers the robustness budget over the whole scheduling horizon, increasing the overall flexibility of the problem, and simplifying the parameter definition of the model. Simulation results show that apart from a more realistic representation of the uncertainties that characterize the problem, the proposed model provides improved results from the feasibility point of view. Moreover, the simplicity and high computational efficiency of the model allow comprehensive evaluations of the results by means of, e.g., parametric sensitivity analyses. Results from such studies indicate that most of the reduction in RVPP profit comes from the ND-RESs energy uncertainty.

Furthermore, for the cases discussed, considering the uncertainty the RVPP prefers to trade energy in the IDMs in which the forecast data is more accurate, resulting in less reserve provided.

The work presented in this paper can be extended in several ways. On one hand, the optimization problem that has been applied to sequential electricity markets can be formulated for the case of joint electricity market clearing. A probabilistic assessment of the proposed robust model will be provided in future works to determine the required uncertainty budget for each uncertain parameter so that the RVPP operator reaches a certain desired income. Moreover, the benefits of RVPP aggregation will be thoroughly evaluated considering different conditions with respect to the case studies discussed in this paper for model validation and testing. A price-maker optimization model of RVPP as a multi-level problem will be provided to address the increasing penetration of RES in the emerging grids. In this context, the regulatory aspects of the system and the estimated strategies of other market competitors have to be taken into account in the developed model. Finally, a multi-level problem to identify the worst realizations of the combination of price and energy uncertainties, i.e., incomes robustness, is also under development.

**Acknowledgements**

This project has received funding from the European Union's Horizon 2020 research and innovation programme under grant agreement No 883985.

**Appendix A. Bertsimas and Sim Robust Approach**

The robust approach of Bertsimas and Sim [22] is summarized in this appendix, which initially considers the linear optimization problem (A.1a)-(A.1c).

$$\max \ \boldsymbol{c'x} \quad (A.1a)$$

$$s.t.:$$

$$\boldsymbol{Ax} \leq \boldsymbol{b} \quad (A.1b)$$

$$\boldsymbol{l} \leq \boldsymbol{x} \leq \boldsymbol{u} \quad (A.1c)$$

In (A.1a), $\boldsymbol{x}$ is the vector of free decision variables; $\boldsymbol{c'}$ is the transposed vector of fixed parameters of the objective function; $\boldsymbol{A}$ is the matrix of uncertain parameters; $a_{ij}$ are the elements of matrix $\boldsymbol{A}$; and it is assumed that $J_i$ is the set of elements in row $i$ of matrix $\boldsymbol{A}$ that are subject to uncertainty. Each entry of $a_{ij}$, $j \in J_i$ is a symmetric bounded random parameter $\tilde{a}_{ij}$ (i.e., $a_{ij} \in [\tilde{a}_{ij} - \hat{a}_{ij}, \tilde{a}_{ij} + \hat{a}_{ij}]$); $\boldsymbol{b}$ is the vector of upper bounds of uncertain constraint (A.1b); $\boldsymbol{l}$ and $\boldsymbol{u}$ are the vectors of lower and upper bounds of inequality constraint (A.1c), respectively.

In (A.1a)-(A.1c), it is assumed that uncertainty only affects matrix $\boldsymbol{A}$ in (A.1b) since if the uncertainty of objective function $\boldsymbol{c}$ needs to be modeled, it can be done by including the constraint $z - \boldsymbol{c'x} \leq 0$ into constraint (A.1b) and maximize the auxiliary objective function $z$.

Then, the non-linear formulation (A.2a)-(A.2e) is proposed as the robust optimization of the problem (A.1a)-(A.1c).

$$\max \ \boldsymbol{c'x} \quad (A.2a)$$

s.t.:

$$\sum_j \tilde{a}_{ij}x_j + \max_{\{S_i \cup \{h_i\}|S_i \subseteq J_i, \lfloor \Gamma_i \rfloor = |S_i|, h_i \in J_i \setminus S_i\}} \left\{ \sum_{t \in S_i} \hat{a}_{ij}y_j + (\Gamma_i - \lfloor \Gamma_i \rfloor)\hat{a}_{ih_i}y_t \right\} \leq b_i \qquad \forall i \qquad \text{(A.2b)}$$

$$-y_j \leq x_j \leq y_j \qquad \forall j \qquad \text{(A.2c)}$$

$$l \leq x \leq u \qquad \text{(A.2d)}$$

$$y \geq 0 \qquad \text{(A.2e)}$$

where set $S_i$ includes the optimization variables in problem (A.2b)-(A.2e) and is defined through a subset of the uncertainty set $J_i$; index $h_i$ contains one optimization variable in problem (A.2b)-(A.2e) and belongs to the uncertainty set $J_i$, while it is not contained in set $S_i$; parameter $\Gamma_i$ takes value in the interval $[0,|J_i|]$ and adjusts the level of robustness that the user chooses for uncertainty.

It is proved that the objective function of the linear problem (A.3a)-(A.3c) is equivalent to the selection of the subset $\{S_i \cup \{h_i\}|S_i \subseteq J_i, \lfloor \Gamma_i \rfloor = |S_i|, h_i \in J_i \setminus S_i\}$ in (A.2b).

$$\sum_{j \in J_i} \hat{a}_{ij}|x_j^*|z_{ij} \qquad \text{(A.3a)}$$

s.t.:

$$\sum_{j \in J_i} z_{ij} \leq \Gamma_i \qquad \text{(A.3b)}$$

$$0 \leq z_{ij} \leq 1 \qquad \forall j \in J_i \qquad \text{(A.3c)}$$

where $x_j^*$ is the protection function of the $i$th constraint; and $z_{ij}$ is a positive variable less than $1$.

Finally, by applying the strong duality theorem to (A.3a)-(A.3c) and substituting the result to the problem (A.2b)-(A.2e), the equivalent linear formulation is obtained as (A.4a)-(A.4h):

$$\max c'x \qquad \text{(A.4a)}$$

s.t.:

$$\sum_j a_{ij}x_j + z_i\Gamma_i + \sum_{j \in J_i} p_{ij} \leq b_i \qquad \forall i \qquad \text{(A.4b)}$$

$$z_i + p_{ij} \geq \hat{a}_{ij}y_j \qquad \forall i, j \in J_i \qquad \text{(A.4c)}$$

$$-y_j \leq x_j \leq y_j \qquad \forall i \qquad \text{(A.4d)}$$

$$l_j \leq x_j \leq u_j \qquad \forall j \qquad \text{(A.4e)}$$

$$p_{ij} \geq 0 \qquad \forall i, j \in J_i \qquad \text{(A.4f)}$$

$$y_j \geq 0 \qquad \forall j \qquad \text{(A.4g)}$$

$$z_i \geq 0 \qquad \forall i \qquad \text{(A.4h)}$$

where $z_i$ and $p_{ij}$ are dual variables related to the parameter uncertainty; $y_j$ is the auxiliary variable that calculates the $x_j$ absolute value function; $l_j$ and $u_j$ are the elements of vectors $l$ and $u$, respectively.

The proposed model by Bertsimas and Sim in (A.4a)-(A.4h) is a linear optimization problem; however, it is shown in [22] that the proposed approach is also valid when the original problem is an MILP (i.e., when some of the variables in the vector $x$ are integer), such as the problem proposed in this paper.

**Appendix B. Sub-region Network Constraints**

In this Appendix, the internal sub-region network of the RVPP, provided that the units in the portfolio are "electrically close" to each other conforming to a defined geographical sub-region, is formulated. The equations for a DC power flow used for active power trading in DAM and SRM are presented below. Note that, for IDMs participation, the time periods vary from $\forall t$ to $\forall t \geq \tau$. The active power flow through transmission line $l$ is represented in (B1.a), whereas the bounds of active power flow through transmission line $l$ are imposed by (B1.b). Finally, the voltage angle reference and angle limits are set by (B1.c) and (B1.d), respectively.

$$p_{lt} = (1/X_l)(\delta_{i(l),t} - \delta_{j(l),t}) \qquad \forall l,t \qquad (B1.a)$$

$$-\overline{P}_l \leq p_{lt} \leq \overline{P}_l \qquad \forall l,t \qquad (B1.b)$$

$$\delta_{b,t} = 0 \qquad b: ref., \forall t \qquad (B1.c)$$

$$-\pi \leq \delta_{b,t} \leq \pi \qquad \forall b,t \qquad (B1.d)$$

In the above constraints, $p_{lt}$ is the active power flow through transmission line $l$ in time period $t$. $X_l$ is the reactance of transmission line $l$. $\delta_{b,t}$ is the voltage angle at bus $b$ in time period $t$. $i(l)$ and $j(l)$ are sending and receiving bus of transmission line $l$, respectively. $\overline{P}_l$ is the capacity of transmission line $l$.

To implement the sub-region network constraints, the supply-demand balancing constraints for each market session defined in Section 4.3.1 are modified according to constraints (B2)-(B6). Constraints (B2.a) and (B2.b) enforce the supply-demand balancing for the RVPP units considering both energy and up/down SR in the DAM+SRM for the main sub-region buses and other buses, respectively.

$$\sum_{r \in R}\left(p_{r,t}^{DA}+r_{r,t}^{SR}\right) + \sum_{\theta \in \Theta}\left(p_{\theta,t}^{DA}+r_{\theta,t}^{SR}\right) - \sum_{l|i(l)=b} p_{lt} + \sum_{l|j(l)=b} p_{lt} = p_t^{DA}+r_t^{SR}+\sum_{d \in D}\left(p_{d,t}^{DA}-r_{d,t}^{SR}\right) \qquad \forall b \in B_m, \forall t \qquad (B2.a)$$

$$\sum_{r \in R}\left(p_{r,t}^{DA}+r_{r,t}^{SR}\right) + \sum_{\theta \in \Theta}\left(p_{\theta,t}^{DA}+r_{\theta,t}^{SR}\right) - \sum_{l|i(l)=b} p_{lt} + \sum_{l|j(l)=b} p_{lt} = \sum_{d \in D}\left(p_{d,t}^{DA}-r_{d,t}^{SR}\right) \qquad \forall b \in B/B_m, \forall t \qquad (B2.b)$$

The supply-demand balancing constraints in the DAM+SRM for the main sub-region buses and other buses are formulated in (B3.a) and (B3.b), respectively.

$$\sum_{r \in R}\left(p_{r,t}^{DA*}+p_{(k=1),r,t}^{ID}+r_{r,t}^{SR}\right) + \sum_{\theta \in \Theta}\left(p_{\theta,t}^{DA*}+p_{(k=1),\theta,t}^{ID}+r_{\theta,t}^{SR}\right)$$
$$- \sum_{l|i(l)=b} p_{lt} + \sum_{l|j(l)=b} p_{lt} = p_t^{DA*}+p_{(k=1),t}^{ID}+r_t^{SR}+\sum_{d \in D}\left(p_{d,t}^{DA*}+p_{(k=1),d,t}^{ID}-r_{d,t}^{SR}\right) \qquad \forall b \in B_m, \forall t \qquad (B3.a)$$

$$\sum_{r \in R} \left( p_{r,t}^{DA*} + p_{(k=1),r,t}^{ID} + r_{r,t}^{SR} \right) + \sum_{\theta \in \Theta} \left( p_{\theta,t}^{DA*} + p_{(k=1),\theta,t}^{ID} + r_{\theta,t}^{SR} \right)$$

$$- \sum_{l|i(l)=b} p_{lt} + \sum_{l|j(l)=b} p_{lt} = \sum_{d \in D} \left( p_{d,t}^{DA*} + p_{(k=1),d,t}^{ID} - r_{d,t}^{SR} \right) \quad \forall b \in B/B_m, \forall t \quad \text{(B3.b)}$$

The supply-demand balancing constraints in the IDMs for the main sub-region buses and other buses are formulated in (B4.a) and (B4.b), respectively.

$$\sum_{r \in R} \left( p_{r,t}^{DA*} + r_{r,t}^{SR} + \sum_{k=1}^{k-1} p_{k,r,t}^{ID*} + p_{k,r,t}^{ID} \right) + \sum_{\theta \in \Theta} \left( p_{\theta,t}^{DA*} + r_{\theta,t}^{SR} + \sum_{k=1}^{k-1} p_{k,\theta,t}^{ID*} + p_{k,\theta,t}^{ID} \right)$$

$$- \sum_{l|i(l)=b} p_{lt} + \sum_{l|j(l)=b} p_{lt} = p_t^{DA*} + r_t^{SR*} + \sum_{k=1}^{k-1} p_{k,t}^{ID*}$$

$$+ p_{k,t}^{ID} + \sum_{d \in D} \left( p_{d,t}^{DA*} - r_{d,t}^{SR} + \sum_{k=1}^{k-1} p_{k,d,t}^{ID*} + p_{k,d,t}^{ID} \right) \quad \forall b \in B_m, \forall k, t \geq \tau \quad \text{(B4.a)}$$

$$\sum_{r \in R} \left( p_{r,t}^{DA*} + r_{r,t}^{SR} + \sum_{k=1}^{k-1} p_{k,r,t}^{ID*} + p_{k,r,t}^{ID} \right) + \sum_{\theta \in \Theta} \left( p_{\theta,t}^{DA*} + r_{\theta,t}^{SR} + \sum_{k=1}^{k-1} p_{k,\theta,t}^{ID*} + p_{k,\theta,t}^{ID} \right)$$

$$- \sum_{l|i(l)=b} p_{lt} + \sum_{l|j(l)=b} p_{lt} = \sum_{d \in D} \left( p_{d,t}^{DA*} - r_{d,t}^{SR} + \sum_{k=1}^{k-1} p_{k,d,t}^{ID*} + p_{k,d,t}^{ID} \right) \quad \forall b \in B/B_m, \forall k, t \geq \tau \quad \text{(B4.b)}$$

**Appendix C. Out-of-sample Assessment**

In this Appendix, an out-of-sample assessment is presented to evaluate the performance of the proposed model compared to [23]. To reach this goal, a set of $|\omega|$ random scenarios with equal probability is generated for DAM and SRM electricity prices, available ND-RESs power production, thermal power output of the solar field of STUs, and demands. The corresponding parameters for each scenario are $\lambda_{t,\omega}^{DA}$, $\lambda_{t,\omega}^{SR,\uparrow}$, $\lambda_{t,\omega}^{SR,\downarrow}$, $P_{r,t,\omega}^{DA}$, $P_{\theta,t,\omega}^{DA}$, and $P_{d,t,\omega}^{DA}$, respectively. The above parameters are assumed to only take the left or right deviated values or, the median value in their corresponding bounds. Then the scenario-based problem (C1) is solved by using the generated scenarios for uncertain parameters and by fixing the obtained market bid results of solving the proposed model in (7), i.e. the energy and reserve bid variables $p_t^{DA*}$, $r_t^{SR,\uparrow*}$, $r_t^{SR,\downarrow*}$, as well as the variable of selection of load profiles $u_{d,p}^*$.

$$\max_{\Xi^{DA+SR}} 1/|\omega| \left\{ \sum_{\omega \in \Omega} \sum_{t \in T} (\lambda_{t,\omega}^{DA} p_t^{DA*} \Delta t + \lambda_{t,\omega}^{SR,\uparrow} r_t^{SR,\uparrow*} + \lambda_{t,\omega}^{SR,\downarrow} r_t^{SR,\downarrow*}) \right.$$

$$\left. - \sum_{\omega \in \Omega} \sum_{t \in T} \sum_{r \in R} C_r^R p_{r,t,\omega}^{DA} \Delta t - \sum_{d \in D} \sum_{p \in P} C_{d,p} u_{d,p}^* - \sum_{\omega \in \Omega} \sum_{t \in T} Z \kappa_{t,w} \Delta t \right\} \quad \text{(C1.a)}$$

s.t.:

$$\sum_{r\in R}\left(p_{r,t,\omega}^{DA}+r_{r,t,\omega}^{SR}\right)+\sum_{\theta\in\Theta}\left(p_{\theta,t,\omega}^{DA}+r_{\theta,t,\omega}^{SR}\right)=p_t^{DA*}+r_t^{SR*}-\kappa_{t,w}+\sum_{d\in D}\left(p_{d,t,\omega}^{DA}-r_{d,t,\omega}^{SR}\right) \quad \forall t,\omega \quad (C1.b)$$

$$\underline{P}_r \leq p_{r,t,\omega}^{DA} - r_{r,t,\omega}^{SR,\downarrow} \quad \forall r,t,\omega \quad (C1.c)$$

$$p_{r,t,\omega}^{DA} + r_{r,t,\omega}^{SR,\uparrow} \leq P_{r,t,\omega}^{DA} \quad \forall r,t,\omega \quad (C1.d)$$

$$0 \leq p_{\theta,t,\omega}^{SF} \leq P_{\theta,t,\omega}^{DA} \quad \forall \theta,t,\omega \quad (C1.e)$$

$$p_{d,t,\omega}^{DA} = P_{d,t,\omega}^{DA} \quad \forall d,t,\omega \quad (C1.f)$$

$$r_{d,t,\omega}^{SR,\uparrow} \leq \underline{\beta}_{d,t} \sum_{p\in P} \widehat{P}_{d,p,t}^{DA} u_{d,p}^* \quad \forall d,t,\omega \quad (C1.g)$$

$$r_{d,t,\omega}^{SR,\uparrow} \leq p_{d,t,\omega}^{DA} - \underline{P}_d \quad \forall d,t,\omega \quad (C1.h)$$

$$r_{d,t,\omega}^{SR,\downarrow} \leq \overline{\beta}_{d,t} \sum_{p\in P} \widehat{P}_{d,p,t}^{DA} u_{d,p}^* \quad \forall d,t,\omega \quad (C1.i)$$

$$r_{d,t,\omega}^{SR,\downarrow} \leq \overline{P}_d - p_{d,t,\omega}^{DA} \quad \forall d,t,\omega \quad (C1.j)$$

$$\left(p_{d,t,\omega}^{DA} + r_{d,t,\omega}^{SR,\downarrow}\right) - \left(p_{d,(t-1),\omega}^{DA} - r_{d,(t-1),\omega}^{SR,\uparrow}\right) \leq \overline{R}_d \Delta t \quad \forall d,t,\omega \quad (C1.k)$$

$$\left(p_{d,(t-1),\omega}^{DA} + r_{d,(t-1),\omega}^{SR,\downarrow}\right) - \left(p_{d,t,\omega}^{DA} - r_{d,t,\omega}^{SR,\uparrow}\right) \leq \underline{R}_d \Delta t \quad \forall d,t,\omega \quad (C1.l)$$

$$r_{d,t,\omega}^{SR,\uparrow} \leq T^{SR} \underline{R}_d^{SR} \quad \forall d,t,\omega \quad (C1.m)$$

$$r_{d,t,\omega}^{SR,\downarrow} \leq T^{SR} \overline{R}_d^{SR} \quad \forall d,t,\omega \quad (C1.n)$$

$$\underline{E}_d \leq \sum_{t\in T}\left(p_{d,t,\omega}^{DA}\Delta t - r_{d,t,\omega}^{SR,\uparrow}\right) \quad \forall d,\omega \quad (C1.o)$$

In problem (C1), the variables $p_{r,t,\omega}^{DA}$, $r_{r,t,\omega}^{SR,\uparrow}$, $r_{r,t,\omega}^{SR,\downarrow}$, $r_{r,t,\omega}^{SR}$, $p_{\theta,t,\omega}^{DA}$, $p_{d,t,\omega}^{DA}$, $r_{d,t,\omega}^{SR,\uparrow}$, $r_{d,t,\omega}^{SR,\downarrow}$, and $r_{d,t,\omega}^{SR}$ related to power and reserve provided by ND-RESs, STUs, and demands are written for each scenario $\omega$. The first three terms in the objective function (C1.a) calculate the operating profit of RVPP in all scenarios for the DAM energy and SRM reserve participation, named by $\Pi^{av}$. The fourth term of (C1.a) determines the average penalization cost due to not providing some or whole part of energy bid in the market, named by $K^{av}$. The net profit of RVPP can be calculated as $\Pi^{av} - K^{av}$. The slack variable $\kappa_{t,\omega}$ is added in the supply-demand balancing constraint (C1.b) and is penalized by parameter $Z$ in the objective function. Constraints (C1.c)-(C1.d) are scenario-based constraints for ND-RESs. Constraint (C1.e) is scenario-based constraint for STUs. Constraints (C1.f)-(C1.o) are scenario-based constraints for demands.

**References**

[1] L. Baringo and M. Rahimiyan, *Virtual power plants and electricity markets*. Springer International Publishing, 2020.
[2] Á. Ortega *et al.*, "Modeling of VPPs for their optimal operation and configuration," POSYTYF Consortium, Tech. Rep., 2021, deliverable 5.1. [Online]. Available: www.posytyf-h2020.eu.
[3] L. A. Roald, D. Pozo, A. Papavasiliou, D. K. Molzahn, J. Kazempour, and A. Conejo, "Power systems optimization under uncertainty: A review of methods and applications," *Electric Power Systems Research*, vol. 214, p. 108725, Jan. 2023.

[4] S. Yu, F. Fang, Y. Liu, and J. Liu, "Uncertainties of virtual power plant: Problems and countermeasures," *Applied Energy*, vol. 239. Elsevier Ltd, pp. 454–470, Apr. 01, 2019.

[5] J. F. Venegas-Zarama, J. I. Munoz-Hernandez, L. Baringo, P. Diaz-Cachinero, and I. De Domingo-Mondejar, "A review of the evolution and main roles of virtual power plants as key stakeholders in power systems," *IEEE Access*, vol. 10. Institute of Electrical and Electronics Engineers Inc., pp. 47937–47964, 2022.

[6] E. Lobato Miguélez, I. Egido Cortés, L. Rouco Rodríguez, and G. López Camino, "An overview of ancillary services in Spain," *Electric Power Systems Research*, vol. 78, no. 3, pp. 515–523, Mar. 2008.

[7] A. Baringo, L. Baringo, and J. M. Arroyo, "Holistic planning of a virtual power plant with a nonconvex operational model: A risk-constrained stochastic approach," *International Journal of Electrical Power & Energy Systems*, vol. 132, p. 107081, 2021.

[8] R. M. Lima, A. J. Conejo, S. Langodan, I. Hoteit, and O. M. Knio, "Risk-averse formulations and methods for a virtual power plant," *Computers and Operations Research*, vol. 96, pp. 350–373, Aug. 2018.

[9] X. Kong, J. Xiao, D. Liu, J. Wu, C. Wang, and Y. Shen, "Robust stochastic optimal dispatching method of multi-energy virtual power plant considering multiple uncertainties," *Applied Energy*, vol. 279, Dec. 2020.

[10] L. Ju, R. Zhao, Q. Tan, Y. Lu, Q. Tan, and W. Wang, "A multi-objective robust scheduling model and solution algorithm for a novel virtual power plant connected with power-to-gas and gas storage tank considering uncertainty and demand response," *Applied Energy*, vol. 250, pp. 1336–1355, Sep. 2019.

[11] D. Wozabal and G. Rameseder, "Optimal bidding of a virtual power plant on the Spanish day-ahead and intraday market for electricity," *European Journal of Operational Research*, vol. 280, no. 2, pp. 639–655, Jan. 2020.

[12] H. Wang, S. Riaz, and P. Mancarella, "Integrated techno-economic modeling, flexibility analysis, and business case assessment of an urban virtual power plant with multi-market co-optimization," *Applied Energy*, vol. 259, p. 114142, Feb. 2020.

[13] H. Nezamabadi and M. Setayesh Nazar, "Arbitrage strategy of virtual power plants in energy, spinning reserve and reactive power markets," *IET Generation, Transmission and Distribution*, vol. 10, no. 3, pp. 750–763, Feb. 2016.

[14] E. Mashhour and S. M. Moghaddas-Tafreshi, "Bidding strategy of virtual power plant for participating in energy and spinning reserve markets-Part I: Problem formulation," *IEEE Transactions on Power Systems*, vol. 26, no. 2, pp. 949–956, May 2011.

[15] L. Ju, Q. Tan, Y. Lu, Z. Tan, Y. Zhang, and Q. Tan, "A CVaR-robust-based multi-objective optimization model and three-stage solution algorithm for a virtual power plant considering uncertainties and carbon emission allowances," *International Journal of Electrical Power and Energy Systems*, vol. 107, pp. 628–643, May 2019.

[16] Y. Zhang, F. Liu, Z. Wang, Y. Su, W. Wang, and S. Feng, "Robust scheduling of virtual power plant under exogenous and endogenous uncertainties," *IEEE Transactions on Power Systems*, 2021.

[17] V. Singh, T. Moger, and D. Jena, "Uncertainty handling techniques in power systems: A critical review," *Electric Power Systems Research*, vol. 203. Elsevier Ltd, Feb. 01, 2022.

[18] A. J. Conejo, M. Carrión, and J. M. Morales, *Decision making under uncertainty in electricity markets*, vol. 153. Boston, MA: Springer US, 2010.

[19] P. Shinde, I. Kouveliotis-Lysikatos, and M. Amelin, "Multistage stochastic programming for VPP trading in continuous intraday electricity markets," *IEEE Transactions on Sustainable Energy*, vol. 13, no. 2, pp. 1037–1048, Apr. 2022.

[20] M. Vahedipour-Dahraie, H. Rashidizadeh-Kermani, M. Shafie-Khah, and J. P. S. Catalão, "Risk-averse optimal energy and reserve scheduling for virtual power plants incorporating demand response programs," *IEEE Transactions on Smart Grid*, vol. 12, no. 2, pp. 1405–1415, Mar. 2021.

[21] A. G. Zamani, A. Zakariazadeh, and S. Jadid, "Day-ahead resource scheduling of a renewable energy based virtual power plant," *Applied Energy*, vol. 169, pp. 324–340, May 2016.

[22] D. Bertsimas and M. Sim, "The price of robustness," *Operations Research*, vol. 52, no. 1, pp. 35–53, Jan. 2004.

[23] M. Rahimiyan and L. Baringo, "Strategic bidding for a virtual power plant in the day-ahead and real-time markets: A price-taker robust optimization approach," *IEEE Transactions on Power Systems*, vol. 31, no. 4, pp. 2676–2687, Jul. 2016.

[24] A. A. Bafrani, A. Rezazade, and M. Sedighizadeh, "Robust electrical reserve and energy scheduling of power system considering hydro pumped storage units and renewable energy resources," *Journal of Energy Storage*, vol. 54, Oct. 2022.

[25] A. Baringo and L. Baringo, "A stochastic adaptive robust optimization approach for the offering strategy of a virtual power plant," *IEEE Transactions on Power Systems*, vol. 32, no. 5, pp. 3492–3504, Sep. 2017.

[26] A. Baringo, L. Baringo, and J. M. Arroyo, "Day-ahead self-scheduling of a virtual power plant in energy and reserve electricity markets under uncertainty," *IEEE Transactions on Power Systems*, vol. 34, no. 3, pp. 1881–1894, May 2019.

[27] Z. Yuanyuan, Z. Huiru, and L. Bingkang, "Distributionally robust comprehensive declaration strategy of virtual power plant participating in the power market considering flexible ramping product and uncertainties," *Applied Energy*, vol. 343, p. 121133, Aug. 2023.

[28] S. Babaei, C. Zhao, and L. Fan, "A data-driven model of virtual power plants in day-ahead unit commitment," *IEEE Transactions on Power Systems*, vol. 34, no. 6, pp. 5125–5135, Nov. 2019.

[29] S. Yu, F. Fang, and J. Liu, "Flexible operation of a CHP-VPP considering the coordination of supply and demand based on a strengthened distributionally robust optimization," *IET Control Theory & Applications*, vol. 17, no. 16, pp. 2146–2161, Nov. 2023.

[30] H. Liu, J. Qiu, and J. Zhao, "A data-driven scheduling model of virtual power plant using Wasserstein distributionally robust optimization," *International Journal of Electrical Power & Energy Systems*, vol. 137, p. 107801, May 2022.

[31] J.-F. Toubeau, J. Bottieau, F. Vallee, and Z. De Greve, "Deep learning-based multivariate probabilistic forecasting for short-term scheduling in power markets," *IEEE Transactions on Power Systems*, vol. 34, no. 2, pp. 1203–1215, Mar. 2019.

[32] S. Camal, A. Michiorri, and G. Kariniotakis, "Optimal offer of automatic frequency restoration reserve from a combined PV/Wind Virtual Power Plant," *IEEE Transactions on Power Systems*, vol. 33, no. 6, pp. 6155–6170, Nov. 2018.

[33] F. Fang, S. Yu, and X. Xin, "Data-driven-based stochastic robust optimization for a virtual power plant with multiple uncertainties," *IEEE Transactions on Power Systems*, vol. 37, no. 1, pp. 456–466, Jan. 2022.

[34] P. González, J. Villar, C. A. Díaz, and F. A. Campos, "Joint energy and reserve markets: Current implementations and modeling trends," *Electric Power Systems Research*, vol. 109, pp. 101–111, Apr. 2014.

[35] O. Oladimeji, Á. Ortega, L. Sigrist, L. Rouco, P. Sánchez-Martín, and E. Lobato, "Optimal participation of heterogeneous, RES-based virtual power plants in energy markets," *Energies (Basel)*, vol. 15, no. 9, p. 3207, Apr. 2022.

[36] Y. Chen, Y. Niu, M. Du, and J. Wang, "A two-stage robust optimization model for a virtual power plant considering responsiveness-based electric vehicle aggregation," *Journal of Cleaner Production*, p. 136690, Mar. 2023.

[37] O. Oladimeji, A. Ortega, L. Sigrist, P. Sanchez-Martin, E. Lobato, and L. Rouco, "Modeling demand flexibility of RES-based virtual power plants," in *2022 IEEE Power & Energy Society General Meeting (PESGM)*, Jul. 2022, pp. 1–5.

[38] I. Llorente García, J. L. Álvarez, and D. Blanco, "Performance model for parabolic trough solar thermal power plants with thermal storage: Comparison to operating plant data," *Solar Energy*, vol. 85, no. 10, pp. 2443–2460, Oct. 2011.

[39] "Red Eléctrica de España, Esios, Markets and prices, Available: https://www.esios.ree.es/es?locale=en."